\documentclass[aps,prb,amsmath,amssymb,superscriptaddress,10pt,a4paper,floatfix,twocolumn]{revtex4-1}

\usepackage{braket}
\usepackage{graphicx}
\usepackage[english]{babel}
\usepackage{color}   
\usepackage{bbm}
\usepackage{tabularx}
\usepackage{bm}
\usepackage{tikz}
\usepackage{lipsum}
\usepackage[pdfpagelabels,plainpages=false,bookmarks=true,colorlinks,linkcolor=red,urlcolor=blue,citecolor=blue]{hyperref}

\newcommand{\unity}{\openone}
\newcommand{\rme}{{\rm e}}
\newcommand{\rmi}{{\rm i}}
\newcommand{\s}{\sigma}

\newcommand{\Fig}[1]{Fig.~\ref{#1}}
\newcommand{\Sec}[1]{Sec.~\ref{#1}}
\newcommand{\App}[1]{Appendix~\ref{#1}}
\newcommand{\Tab}[1]{Table~\ref{#1}}
\newcommand{\Ref}[1]{Ref.~\onlinecite{#1}}

\renewcommand{\O}{\mathcal{O}}

\newcommand{\rbra}[1]{(#1|}
\newcommand{\rket}[1]{|#1)}

\def\figpath{.}
\def\msuc{multi site unit cell}

\newcommand{\xqn}{q}
\newcommand{\gsqn}{\tilde{q}}
\newcommand{\mom}{p}
\newcommand{\fac}{\rme^{\rmi \mom N}}
\newcommand{\cfac}{\rme^{-\rmi \mom N}}

\begin{document}
\title{Topological nature of spinons and holons: Elementary excitations from matrix product states with conserved symmetries}

\author{V. \surname{Zauner-Stauber}}
\affiliation{Vienna Center for Quantum Technology, University of Vienna, Boltzmanngasse 5, 1090 Wien, Austria}
\author{L. \surname{Vanderstraeten}}
\affiliation{Ghent University, Faculty of Physics, Krijgslaan 281, 9000 Gent, Belgium}
\author{J. \surname{Haegeman}}
\affiliation{Ghent University, Faculty of Physics, Krijgslaan 281, 9000 Gent, Belgium}
\author{I.P. \surname{McCulloch}}
\affiliation{ARC Centre of Excellence for Engineered Quantum Systems, School of Mathematics and Physics,
The University of Queensland, St Lucia, QLD 4072, Australia}
\author{F. \surname{Verstraete}}
\affiliation{Vienna Center for Quantum Technology, University of Vienna, Boltzmanngasse 5, 1090 Wien, Austria}
\affiliation{Ghent University, Faculty of Physics, Krijgslaan 281, 9000 Gent, Belgium}
 
\begin{abstract}
We develop variational matrix product state (MPS) methods with symmetries to determine dispersion relations of one dimensional quantum lattices as a function of momentum and preset quantum number. We test our methods on the XXZ spin chain, the Hubbard model and a non-integrable extended Hubbard model, and determine the excitation spectra with a precision similar to the one of the ground state. The formulation in terms of quantum numbers makes the topological nature of spinons and holons very explicit. In addition, the method also enables an easy and efficient direct calculation of the necessary magnetic field or chemical potential required for a certain ground state magnetization or particle density.
\end{abstract}
\maketitle

\section{Introduction}
\label{sec:intro}
Matrix product state (MPS)\cite{MPS1_FNW,MPS4_PVWC,MPS5_VMC,MPS6_S} based methods such as DMRG,\cite{DMRG1,DMRG2,DMRG_MC} TEBD\cite{TEBD} and VUMPS\cite{VUMPS} have proven to be invaluable tools for simulating ground states of one dimensional quantum lattice models. By formulating those MPS methods in terms of manifolds and tangent spaces,\cite{MPSTP} it has recently been shown that excitation spectra or dispersion relations as a function of momenta can readily be determined once the ground state is written in terms of a uniform (translation invariant) ground state.\cite{MPS_Excitations_variational,MPS_excitations} Those tangent space methods extend the works of \"Ostlund and Rommer,\cite{OstlundRommer} in which a slightly more limited ansatz was used. In this paper, we extend those tangent space methods to accommodate for $U(1)$  symmetries, which are necessary to simulate quantum systems exhibiting a large amount of entanglement to good precision, such as e.g. the Hubbard model. The symmetric formulation further allows for targeting excitations with certain quantum numbers only, which greatly helps in disentangling rich excitation spectra of models which e.g. host several different types of elementary excitations.

In \Sec{sec:sumps} we develop the theory of symmetric uniform MPS, while in \Sec{sec:sumps_excitations} we introduce the necessary tools for formulating the excitation ansatz in the presence of symmetries, where we also generalize to multi site unit cells. This is done both for topologically trivial and nontrivial excitations which are domain wall like, such as spinons and holons. In \Sec{sec:results}, we demonstrate the usefulness of the methods by simulating excitation spectra of the integrable XXZ and Fermi Hubbard model, as well as the non-integrable extended Fermi Hubbard model. The use of symmetries makes the topologically nontrivial nature of spinons and holons very clear and intuitive. Finally, we conclude with a summary and outlook in \Sec{sec:conclusion}.

\section{Symmetric Uniform MPS}
\label{sec:sumps}

We begin by defining properties for symmetric uniform MPS (suMPS), 
where we restrict the discussion to the case of abelian symmetries (e.g. $\mathbb{Z}_{n}$ parity, or $U(1)$ like particle number or magnetization).
While symmetric tensor networks and the use of conserved quantities in tensor network algorithms have been addressed in numerous previous works \cite{OstlundRommerIMPS, Syms_DPG, DMRG_MC, IDMRG, ITEBD_Hastings, Syms_Sukhi}, we reiterate here in detail their consistent use in the context of MPS in the thermodynamic limit.

In the following we closely use and follow notation and nomenclature of \Ref{VUMPS} and restate here only the most important concepts. For details we refer the reader to \Ref{VUMPS} (in particular Sec. II.A and II.E.). We consider a translation invariant uniform MPS in the thermodynamic limit in the mixed canonical representation
\begin{subequations}
\label{eq:psi_mixed}
\begin{align}
\ket{\Psi(A)}&=\sum_{\bm{\s}}(\ldots  A_{L}^{\s_{n-1}} A_{C}^{\s_{n}} A_{R}^{\s_{n+1}} \ldots)\ket{\bm{\s}}\label{eq:psiAC}\\
 &=\sum_{\bm{\s}}(\ldots  A_{L}^{\s_{n-1}}A_{L}^{\s_{n}} C A_{R}^{\s_{n+1}} A_{R}^{\s_{n+2}} \ldots)\ket{\bm{\s}}.\label{eq:psiC}
\end{align} 
\end{subequations}
Here $A_{L}, A_{R} \in\mathbbm{C}^{D\times d\times D}$, with $d$ the local Hilbert space dimension and $D$ the MPS bond dimension, both describe the same state in different gauge representations and are related by the gauge transformation matrix $C$ via
\begin{equation}
A_{L}^{\s} C = C A_{R}^{\s}= A_{C}^{\s},
 \label{eq:AC}
\end{equation} 
where we have also defined the center site matrix $A_{C}$. $A_{L}$ and $A_{R}$ then fulfill the left and right gauge constraints
\begin{subequations}
\label{eq:gauges}
\begin{align}
 \sum_{\s}{A_{L}^{\s}}^{\dagger}A_{L}^{\s}&=\unity &   \sum_{\s}A_{L}^{\s}\,CC^{\dagger}\, {A_{L}^{\s}}^{\dagger}&=CC^{\dagger}\label{eq:leftgauge}\\
 \sum_{\s}A_{R}^{\s}{A_{R}^{\s}}^{\dagger}&=\unity &   \sum_{\s}{A_{R}^{\s}}^{\dagger}\,C^{\dagger}C\, A_{R}^{\s}&=C^{\dagger}C.\label{eq:rightgauge}
\end{align}
\end{subequations}
and the singular values of $C$ correspond to the Schmidt values of a bipartition of the state $\ket{\Psi(A)}$.

An $N$ site unit cell uMPS, which is invariant under a translation over $N$ sites, is described by $N$ independent MPS matrices ${A{(k)}}, k=1,\ldots,N$, which define the unit cell tensor
\begin{equation}
 \mathbb{A}^{\Sigma_{n}}=A(1)^{\s_{nN+1}}\ldots A(N)^{\s_{nN+N}},
\end{equation} 
where $\Sigma_{n}=(\s_{nN + 1},\ldots,\s_{nN + N})$ is a combined index. For an $N$ site unit cell uMPS we then write
\begin{equation}
 \ket{\Psi(\mathbb{A})}=\sum_{\bm{\s}}(\ldots  \mathbb{A}^{\Sigma_{n-1}}\mathbb{A}^{\Sigma_{n}} \mathbb{A}^{\Sigma_{n+1}} \ldots)\ket{\bm{\s}},
\end{equation} 
where the integer $n$ labels unit cells, not sites. Here we have not explicitly specified the gauge representations of the individual MPS matrices within the unit cell, but we will henceforth assume all states to be in the mixed canonical representation \eqref{eq:psi_mixed}.

Global symmetries of an MPS that is invariant under certain symmetry operations can be easily encoded in symmetry properties of the local MPS matrices $A$. For MPS with abelian symmetries this simply amounts to attaching quantum numbers to all indices appearing in tensor contractions and constraining the MPS matrices $A$ to transform as irreducible representations of the global symmetry group. 
In practice, states on the physical and virtual level are therefore grouped into distinct quantum number sectors, and the matrices are of sparse block form. The action of the symmetry group then determines which combinations of quantum numbers are allowed, i.e. which blocks are non zero. We denote such symmetric MPS matrices as 
\begin{align}
&A^{(s,\s)}_{(a,\alpha)(b,\beta)}& b&=a\oplus s
\label{eq:smps}
\end{align} 
Here, $a$ and $b$ are quantum numbers labeling symmetry sectors on the virtual level and $\alpha$, $\beta$ label states within these sectors (degeneracy indices), while $s$ denotes the quantum numbers associated with the local physical states $\s$, determined by a choice of numerical representation $s(\sigma)$ (see below for examples), and $\oplus$ denotes the group action. 
\footnote{More precisely, it is the fusion structure of the irreducible representations, which in the case of abelian symmetries is however isomorphic to the group action.}
In the presence of several symmetries, the quantum numbers are multi valued.
In the following we will often omit quantum number labels or degeneracy indices for better readability. To avoid ambiguity, we denote quantum numbers with Latin letters and physical/degeneracy indices with Greek letters. At times we also write -- in a slight abuse of notation -- $A^{\s}_{ab}$, where it is understood that $b=a\oplus s(\s)$. 

In most cases the group action reduces to a simple (perhaps modular) addition/subtraction of properly defined quantum numbers, which we represent as (tuples of) rational numbers. In the following we will therefore denote $a\oplus b$ just as $a+b$, and the action with the inverse $a\oplus\bar{b}$ as $a-b$, where $b\oplus\bar{b}=0$. Furthermore, we regard $s(\s)$ as a numerical representation of the physical state for quantum number arithmetic only, not necessarily as a strict group representation. 

Without loss of generality, in \eqref{eq:smps} we have implicitly defined $s$ and $a$ as \textit{ingoing}, and $b$ as \textit{outgoing} quantum numbers.
\footnote{It is worth noting that the symmetric MPS matrices so defined behave like the transpose of the corresponding symmetry operator: e.g. for the $S=1/2$ ladder operator $S^{+}_{ab}$ we would have $a=b+1$. In that regard we could have also defined \eqref{eq:smps} that way, as the definition is just a matter of choice, as long as this choice is consistently followed subsequently.}
Upon concatenating symmetric MPS matrices, outgoing indices are then connected to ingoing indices only, and only sector blocks with matching quantum numbers are contracted. For example, a concatenation of two such matrices then yields
\begin{align}
 C^{\s_{1}\s_{2}}_{ac}&=\sum_{b}A^{\s_{1}}_{ab}B^{\s_{2}}_{bc}& c&=a + s(\s_{1}) + s(\s_{2}).
\end{align} 

Quantum states on finite systems of size $L$ and with a certain quantum number $Q$ can then be constructed by defining the quantum number of the left virtual boundary state to be zero, and the quantum number of the right virtual boundary state to be the desired quantum number $Q$.
\footnote{
More specifically, we attach a single dummy ingoing quantum number $a=0$ to the leftmost, and a single dummy outgoing quantum number $b=Q$ to the rightmost MPS matrix.
}
Using symmetric MPS matrices \eqref{eq:smps}, only basis states $\ket{\s_{1}\ldots \s_{L}}$ fulfilling the constraint $\sum_{j}s(\s_{j})=Q$ will then contribute to the overall state. 
This procedure is good practice and widely used in implementations of symmetry exploiting MPS algorithms.\cite{DMRG_MC, MPS6_S}

However, quantum numbers of ground states with $U(1)$ symmetry in particular usually scale with the system size $L$ (e.g. particle number $N=L/2$ at half filling), and the above scheme is therefore not scalable to the thermodynamic limit. In an equivalent, more homogeneous formulation, one can however demand both left and right virtual boundary vectors to have the same quantum number,
\footnote{In fact, the left and right boundary quantum numbers need not even be zero, they just have to be equal. Virtual $U(1)$ quantum numbers are therefore always defined up to a global offset only.}
and to subtract the value of the desired quantum number \textit{density} $\gsqn=Q/L$ as part of a modified group action for each individual MPS matrix
\begin{equation}
 b=a + s - \gsqn=a + \tilde{s}
 \label{eq:symop_density}
\end{equation} 
Here we have defined a \textit{shifted} quantum number
\begin{equation}
\tilde{s} =s - \gsqn
\label{eq:mod_qn}
\end{equation} 
for the physical index, which is the original quantum number $s$, offset by the desired overall density $\gsqn$.

This scheme is now easily scalable to the thermodynamic limit and we endow uniform MPS matrices 
with this modified group action \eqref{eq:symop_density} to obtain \textit{symmetric uniform MPS} (suMPS) with well defined quantum number \textit{densities}. The generalization to \msuc s is straightforward  by endowing every matrix within the unit cell with the modified group action \eqref{eq:symop_density} and a consistent definition of the quantum number density. Consequently, the unit cell size $N$ has to be chosen in accordance with the desired quantum number density $\gsqn$. For example, a spin $S=1/2$ suMPS with zero magnetization requires $N$ even in order to host an equal number of up and down spins. Choosing a single site unit cell in this case results in a superposition of possible zero magnetization suMPS and hence a non injective MPS (i.e. the transfer matrix has multiple dominant eigenvalues with magnitude one).

\begin{figure*}[ht]
 \centering
 \includegraphics[width=0.9\linewidth,keepaspectratio=true]{\figpath/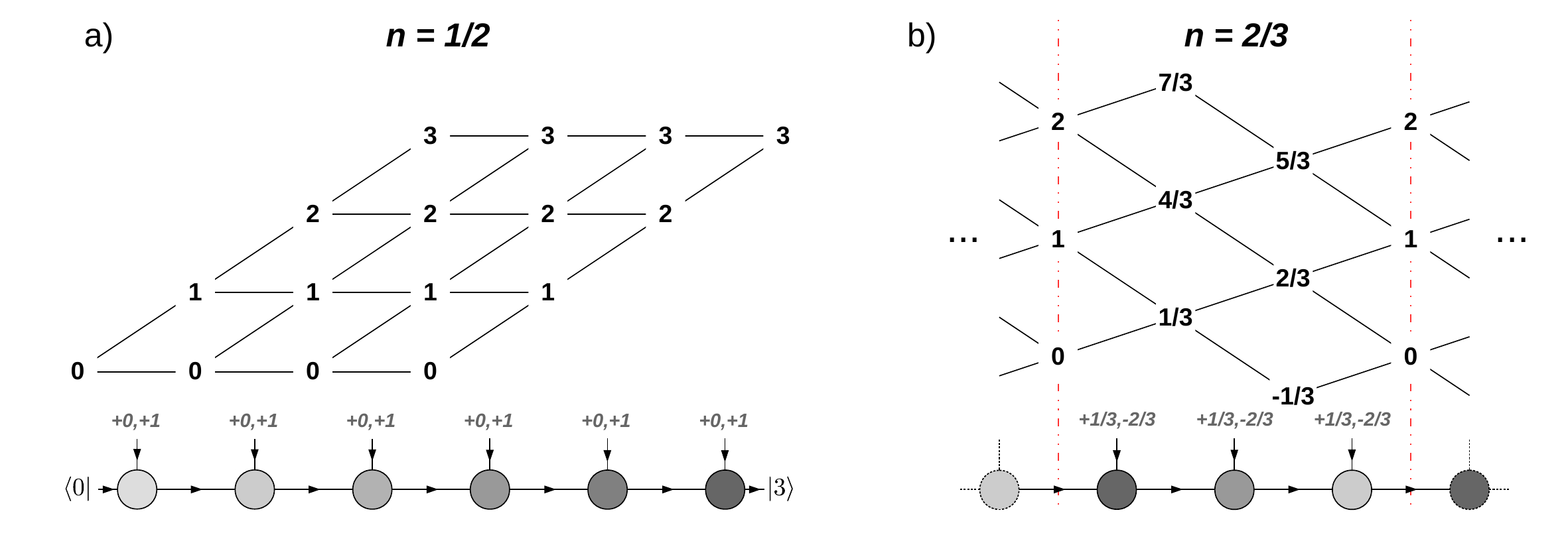}
 \caption{Examples for possible quantum number sectors of MPS on a system of spinless fermions with fixed particle number. The lines represent possible paths, corresponding to valid sequences of available quantum number sectors. (a) Possible quantum number sectors in the conventional finite size representation of a $6$ site system at half filling (i.e. particle density $n=1/2$). Consequently the left and right boundary quantum numbers are $0$ and $3$ respectively, and the on site quantum numbers are $s=0,1$. (b) Possible quantum number sectors for an infinite system at $2/3$ filling and a $3$ site unit cell, with $3$ quantum number sectors per bond. Here, the (shifted) on site quantum numbers are $\tilde{s}=1/3,-2/3$. In typical low energy states, quantum number sectors corresponding to high fluctuations around the desired particle number density will in general be suppressed by small Schmidt values, which will be truncated away in a finite bond dimension MPS approximation, leaving a finite number of remaining sectors, even in infinite systems. Notice also the appearance of negative labels, as quantum numbers here are only defined up to an arbitrary global offset. In both cases we have marked in- and outgoing quantum numbers with arrows.}
 \label{fig:qn_groundstate}
\end{figure*}

We show concrete examples for shifted quantum numbers $\tilde{s}$ and required unit cell sizes $N$ for spin $S=1/2$ states with fixed magnetization densities $m$ in \Tab{tab:qn_xxz}, and states of spinful electrons with fixed particle and magnetization densities $(n,m)$ in \Tab{tab:qn_hub}.
\footnote{
In general, the choice of the physical quantum number labels is not unique.
For example, for spinful electrons, instead of particle and magnetization density one could also choose the densities of up and down spin electrons. In a practical implementation it is furthermore desirable to scale and shift all quantum numbers to e.g. integers to obtain an efficient representation.} 
For an illustration of the conventional finite size scheme and the modified scheme for infinite systems, see \Fig{fig:qn_groundstate}.

It is worth noting that due to the quantum number density offset the shifted quantum numbers in general do not directly correspond to true quantum numbers of the symmetry group. This is because we have distributed the quantum number shift (to achieve a certain quantum number density) homogeneously over the unit cell, instead of applying a single offset at the unit cell boundary. Consequently, the shifted quantum numbers can be interpreted as a combination of the true quantum numbers of the symmetry group combined with quantum numbers of \textit{fractional} applications of the unit cell translation operator (i.e. translations over $n<N$ sites). The shifted quantum numbers therefore also bear information about the location within the unit cell. For example, in the 3 site unit cell suMPS with particle density $n=2/3$ in \Fig{fig:qn_groundstate}, the sets of possible quantum numbers on the three bonds are mutually exclusive, i.e. on the first bond there are only integers, while on the second bond the shifted quantum numbers are integers shifted by $+1/3$, and finally on the third bond they are integers shifted by $+2/3$.

\begin{table}[t]
  \begin{tabular}{c|c|ccc}
  $m$			& $-$		& $0$ 		& $1/6$ 	& $-1/4$\\
  \hline
  $\ket{\uparrow}$	& $1/2$		& $1/2$		& $1/3$ 	& $3/4$ \\
  $\ket{\downarrow}$	& $-1/2$	& $-1/2$ 	& $-2/3$ 	& $-1/4$\\ 
  \hline
  $N$ 			& $-$ 		& $2$ 		& $3$ 		& $4$
 \end{tabular} 
\caption{A few examples of $\tilde{s}(\s)$ and required unit-cell sizes $N$ for various magnetization densities $m$ for a $S=1/2$ chain (with $\ket{\s}=\{\ket{\uparrow},\ket{\downarrow}\}$). The first column denotes the chosen (unmodified) quantum number representations $s(\s)$.}
\label{tab:qn_xxz}
\end{table} 

\begin{table}[t]
 \begin{tabular}{c|c|ccc}
  $(n,m)$ 			& $-$ 		& $(1,0)$ 	& $(2/3,0)$ 	& $(5/4,-1/8)$\\
  \hline
 $\ket{0}$ 			& $(0,0)$ 	& $(-1,0)$ 	& $(-2/3,0)$ 	& $(-5/4,1/8)$ \\
 $\ket{\downarrow}$ 		& $(1,-1/2)$ 	& $(0,-1/2)$ 	& $(1/3,-1/2)$ 	& $(-1/4,-3/8)$ \\
 $\ket{\uparrow}$ 		& $(1,1/2)$ 	& $(0,1/2)$ 	& $(1/3,1/2)$ 	& $(-1/4,5/8)$\\
 $\ket{\downarrow\uparrow}$	& $(2,0)$	& $(1,0)$	& $(4/3,0)$ 	& $(3/4,1/8)$\\
 \hline
 $N$ 				& $-$ 		& $2$ 		& $3$ 		& $4$
 \end{tabular} 
\caption{A few examples of $\tilde{s}(\s)$ and required unit cell size $N$ for various particle number and magnetization densities $(n,m)$ for a chain of spinful electrons (with $\ket{\s}=\{\ket{0},\ket{\downarrow}, \ket{\uparrow}, \ket{\downarrow\uparrow}\}$). The first column denotes the chosen (unmodified) quantum number representations $s(\s)$.}
\label{tab:qn_hub}
\end{table}

\section{Symmetric Variational Ansatz for Elementary Excitations}
\label{sec:sumps_excitations}

We now generalize the variational ansatz for low energy excitations presented in \Ref{MPS_Excitations_variational} to \msuc s and abelian symmetries. As a foundation we start from a variationally optimized suMPS ground state approximation $\ket{\Psi(A)}$ of bond dimension $D$ in mixed canonical representation, where for now we focus on single site unit cells and generalize to \msuc s later.
A suitable suMPS ground state approximation can be obtained from e.g. a symmetric implementation of the algorithm presented in \Ref{VUMPS} using suMPS as described in \Sec{sec:sumps}.

In a slight reformulation of the ansatz in \Ref{MPS_Excitations_variational} we write for a low-energy excitation with momentum $\mom$ in the mixed canonical representation
\begin{equation}
 \ket{\Phi_{\mom}(B)}=\sum_{n,\bm{\s}}\rme^{i\mom n}\left( \ldots A_{L}^{\s_{n-1}}B^{\s_{n}}\tilde{A}_{R}^{\s_{n+1}}\ldots \right)\ket{\bm{\s}}.
 \label{eq:excitation_ansatz}
\end{equation} 
Here, the matrices left and right of $B$ are in left and right canonical representation respectively. Furthermore, for a topologically trivial excitation $A_{L}$ and $\tilde{A}_{R}$ represent the same state, i.e. $|\braket{\Psi(A)|\Psi(\tilde{A})}|=1$. For a topologically nontrivial excitation $A_{L}$ and $\tilde{A}_{R}$ represent different ground states (e.g. within the degenerate ground space in a symmetry broken phase), and the excitation is of domain wall type. The above ansatz only captures elementary excitations and their bound states well,\cite{MPS_Excitations_variational} (or more precisely, isolated excitations branches  \cite{MPS_excitations}) while an accurate representation of scattering states in a continuum requires a more complicated ansatz,\cite{Laurens_Scattering_1,Laurens_Scattering_2} whose symmetric formulation for multi site unit cells we leave for future work. 

In the above mixed canonical representation, the parameterization of the perturbation matrix $B$ in the left and right tangent space gauge\cite{MPS_Excitations_variational} then reduces to
\begin{subequations}
 \label{eq:B_gauges}
\begin{align}
 B^{\s}&=V_{L}^{\s}\,x_{L}, \label{eq:B_left_gauge}\\
 B^{\s}&=x_{R}\,V_{R}^{\s}. \label{eq:B_right_gauge}
\end{align}
\end{subequations}
Here $V_{L}^{\s}$ and $V_{R}^{\s}$ are the left and right null spaces of $A_{L}^{\s}$ and $\tilde{A}_{R}^{\s}$ respectively, i.e. $\sum_{\s}{V_{L}^{\s}}^{\dagger}A_{L}^{\s}=\sum_{\s}\tilde{A}_{R}^{\s}{V_{R}^{\s}}^{\dagger}=0$, and the matrices $x_{L}, x_{R}\in\mathbbm{C}^{D\times (d-1)D}$ contain the $(d-1)D^{2}$ variational degrees of freedom for the ansatz. Without loss of generality we will henceforth use the left tangent space gauge \eqref{eq:B_left_gauge} and drop the subscripts $L/R$ for $V$ and $x$.

Variational approximations of excited states with a certain fixed momentum $\mom$ can then be obtained from solving an effective eigenvalue problem of a (momentum-dependent) effective Hamiltonian defined in the space of these variational parameters\cite{MPS_Excitations_variational}
\begin{equation}
 \mathcal{H}^{\rm eff}_{\mom}\vec{x}^{[j]}=e_{\mom}^{[j]}\vec{x}^{[j]},
 \label{eq:eff_EV}
\end{equation} 
where $j=1,\ldots,(d-1)D^{2}$ and $\vec{x}$ denotes the vectorization of $x$. Notice that in contrast to the original formulation, in the above mixed canonical representation all necessary operations can now be performed without taking any (possibly) ill-conditioned inverses. 

For an $N$ site unit cell ansatz, we again start from a variationally optimized (but here $N$ site unit cell) ground state approximation $\ket{\Psi(\mathbb{A})}$ in mixed canonical representation and introduce local perturbations in the form of local perturbation matrices $B(k)$ on each site. We 
collect all of these single site contributions into one single unit cell perturbation matrix
\begin{equation}
\begin{split}
 \mathbbm{B}^{\Sigma_{n}}=&B(1)^{\s_{nN+1}}\tilde{A}(2)_{R}^{\s_{nN+2}}\ldots \tilde{A}(N)_{R}^{\s_{nN+N}} +\\
 &A(1)_{L}^{\s_{nN+1}}B(2)^{\s_{nN+2}}\ldots \tilde{A}(N)_{R}^{\s_{nN+N}} + \\
 & \ldots +\\
 &A(1)_{L}^{\s_{nN+1}}\ldots B(N-1)^{\s_{nN+N-1}}\tilde{A}(N)_{R}^{\s_{nN+N}}+\\
 &A(1)_{L}^{\s_{nN+1}}\ldots A(N-1)_{L}^{\s_{nN+N-1}}B(N)^{\s_{nN+N}}
\end{split}
\end{equation}
and write the full \msuc\ ansatz with momentum $0\leq p<\frac{2\pi}{N}$ as 
\begin{equation}
\ket{\Phi_{\mom}(\mathbbm{B})}=\sum_{n,\bm{\s}}\rme^{i\mom N n}\left( \ldots \mathbbm{A}^{\Sigma_{n-1}}_{L} \mathbbm{B}^{\Sigma_{n}} \tilde{\mathbbm{A}}^{\Sigma_{n+1}}_{R}\ldots\right)\ket{\bm{\s}}.
\label{eq:msuc_excitation_ansatz}
\end{equation} 
Here again the integer $n$ enumerates unit cells and we parameterize $B^{\s}(k)=V^{\s}(k)\,x(k)$, where $\sum_{\s}{V(k)^{\s}}^{\dagger}A(k)_{L}^{\s}=0$ for $k=1,\ldots,N$.
The variational energy is then a quadratic function of the concatenation of all $N$ parameter vectorizations $\vec{x}=\bigoplus_{k}\vec{x}(k)$, and \msuc\ excited state approximations can be obtained from solving an effective eigenvalue problem of the same type as \eqref{eq:eff_EV}, but with a larger and more complex effective Hamiltonian $\mathcal{H}^{\rm eff}_{\mom}$ (for an explicit construction of $\mathcal{H}^{\rm eff}_{\mom}$ see Appendix~\ref{sec:Heff}). Note that, contrary to blocking sites in e.g. a regular DMRG calculation, here the number of variational parameters scales linearly with $N$, enabling an efficient treatment of large unit cell sizes without sweeping.

By definition and construction of the variational ansatz and $\mathcal{H}^{\rm eff}_{\mom}$, the excitation energies $e^{[j]}_{\mom}$ obtained from \eqref{eq:eff_EV} are (positive) energy differences to the \textit{extensive} ground state energy $E_{0}$. Hence, while $E_{0}$ is of $\O(L)$, the excitation energies $e^{[j]}_{\mom}$ are of $\O(1)$. Likewise, while $U(1)$ quantum numbers -- like particle number or magnetization -- are extensive for ground states, low lying excitations are characterized by quantum number \textit{differences} of $\O(1)$ to the ground state. Popular examples for this are spin flip or few particle excitations.

In the context of the variational ansatz \eqref{eq:excitation_ansatz}, such relative differences of $\O(1)$ can be perfectly well understood as being caused by the single perturbation matrix $B$ on top of the homogeneous, extensive ground state background generated by the MPS matrices $A_{L}$ and $\tilde{A}_{R}$. We can therefore control and fix quantum numbers of excited states through the quantum numbers of $B$.

More specifically, in the parameterization \eqref{eq:B_gauges}, $V_{L}$ and $V_{R}$ necessarily have the same symmetry sectors and quantum number labels as $A_{L}$ and $\tilde{A}_{R}$. We can however attach a non-trivial quantum number $\xqn$ to the matrix $x$ that contains the variational parameters, i.e. $x$ is block (off-)diagonal and we write 
\begin{equation}
 x^{[\xqn]}_{ab},\qquad b=a - \xqn.
\end{equation} 
Note, that $\xqn$ is an (outgoing) quantum number associated with $x$ itself, rather than a physical index. This is very intuitive, as $V$ takes care of the homogeneous ground state density contribution on that site, while the quantum number $\xqn$ of $x$ directly controls the quantum number \textit{difference} of the excitation with respect to the ground state. For $B$ we then have
\begin{equation}
 B^{[\xqn] \s}_{ab}=\sum_{c} V^{\s}_{ac}\,x^{[\xqn]}_{cb},\qquad b = a + \tilde{s}(\s) - \xqn.
\end{equation} 
and we denote the generated symmetric excitation ansatz as $\ket{\Phi^{[\xqn]}_{p}(\mathbbm{B})}$.

From the structure of \eqref{eq:excitation_ansatz} it is clear that the values of $\xqn$ have to be such that the outgoing quantum numbers of $B$ match the ingoing quantum numbers of $\tilde{A}_{R}$. Depending on $A_{L}$ and $\tilde{A}_{R}$, only certain values of $\xqn$ are therefore allowed. 

For example, 
in a system of spinless fermions with $s(\s)=0,1$, a single particle excitation is characterized by $\xqn=1$, while a hole excitation corresponds to $\xqn=-1$. Likewise, 
on a $S=1/2$ spin chain with $s(\s)=\pm 1/2$, single magnons (spin flips) are characterized by $\xqn=\pm 1$. This holds regardless of the ground state particle/magnetization density, which is encoded in $V$. 

The above are typical examples for topologically trivial excitations, where $\tilde{A}_{R}=A_{R}$, i.e. $A_{L}$ and $\tilde{A}_{R}$ have the same quantum number sectors, and the quantum number of the excitation is well defined. The quantum numbers of fractional excitations such as spinons and holons however necessarily require them to be of topologically non trivial nature (see below), where $\tilde{A}_{R}\neq A_{R}$ and the quantum numbers of $A_{L}$ and $\tilde{A}_{R}$ can in principle differ by an arbitrary offset. Just as the momentum $p$ (cf. \Ref{MPS_Excitations_variational}), the quantum number $\xqn$ of a topologically nontrivial excitation therefore seems to be completely arbitrary. Again, this is an artifact of open boundary conditions and fixing this ambiguity depends on the nature of the excitation (see Sec.~\ref{sec:XXZ_results} and \ref{sec:HUB_results}).

\section{Results}
\label{sec:results}

\subsection{Spinons and Magnons in the S=1/2 XXZ Antiferromagnet}
\label{sec:XXZ_results}

\begin{figure*}[t]
 \centering
 \includegraphics[width=0.9\linewidth,keepaspectratio=true]{\figpath/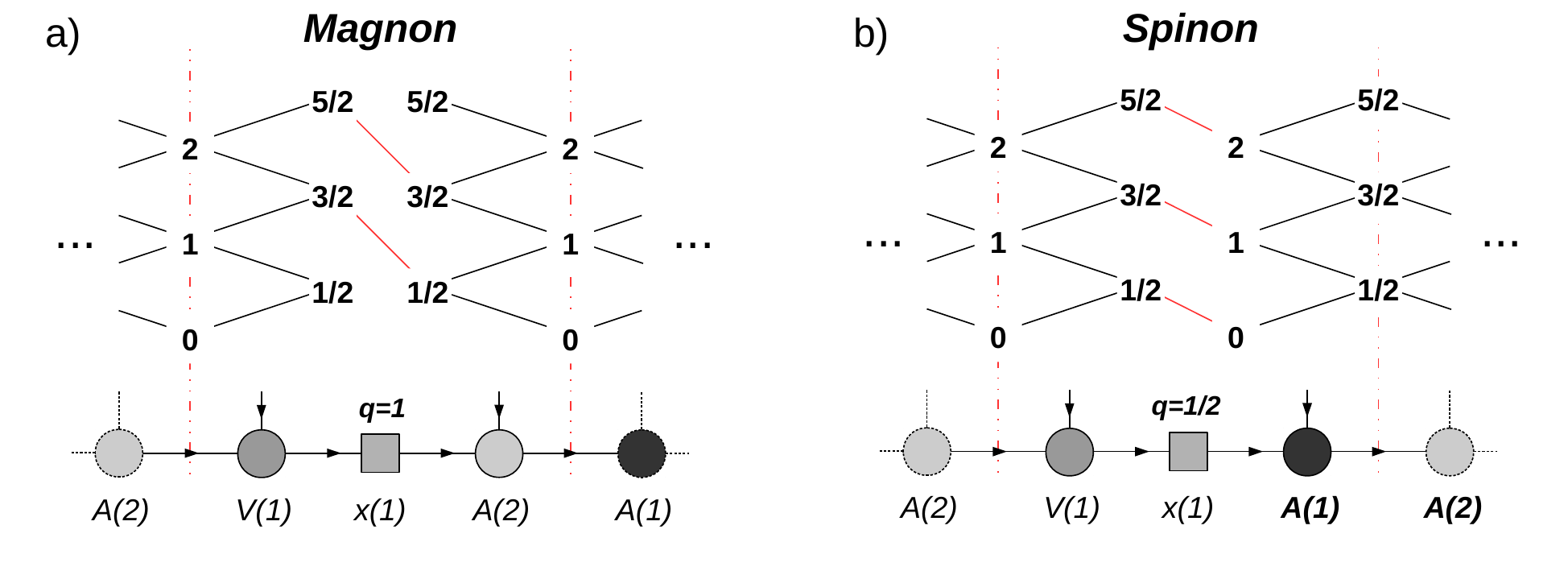}
 \caption{Possible constructions and quantum number sectors for (a) a topologically trivial (magnon) and (b) a topologically nontrivial (spinon) excitation in the antiferromagnetic $S=1/2$ XXZ model. Here we only show one contribution to \eqref{eq:msuc_excitation_ansatz} in a two site unit cell ansatz. Notice the translated state right of $x$ in (b) to enable half integer excitation quantum numbers.}
 \label{fig:qn_excitations}
\end{figure*}
As a first prototypical example we study low energy excitations of the one dimensional $S=1/2$ XXZ model in a magnetic field
\begin{equation}
 H_{\rm XXZ}=\sum_{j}X_{j}X_{j+1} + Y_{j}Y_{j+1} + \Delta Z_{j}Z_{j+1} - h Z_{j} .
 \label{eq:XXZ_Ham}
\end{equation} 
Here $X$, $Y$ and $Z$ are $S=1/2$ spin operators, $\Delta$ is the anisotropy parameter and $h$ is an external magnetic field. The energies for the ground state and low energy excitations in the thermodynamic limit are known exactly.\cite{Takahashi}

\begin{figure*}[t]
 \centering
 \includegraphics[width=0.495\linewidth,keepaspectratio=true]{\figpath/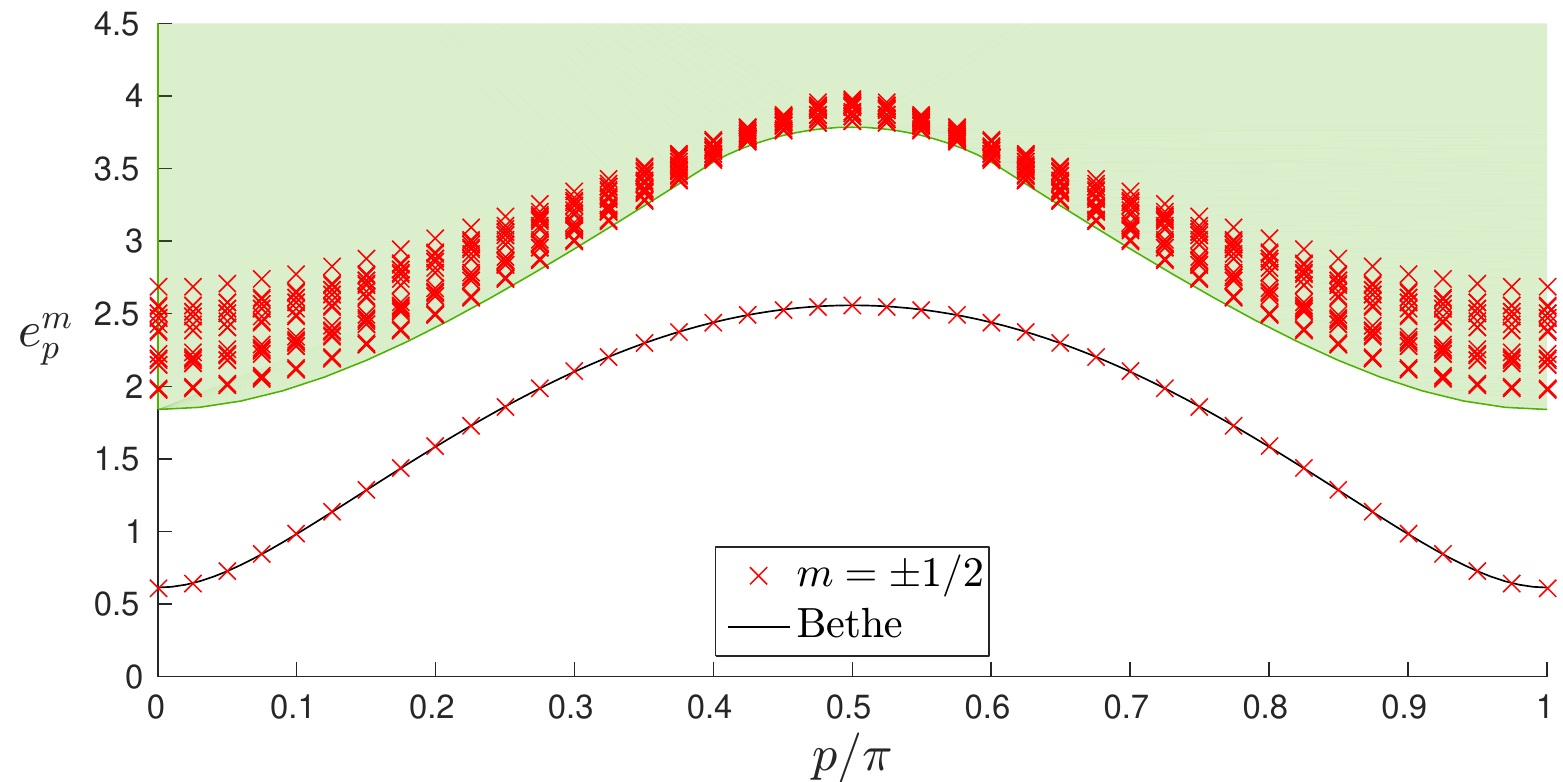}
 \includegraphics[width=0.495\linewidth,keepaspectratio=true]{\figpath/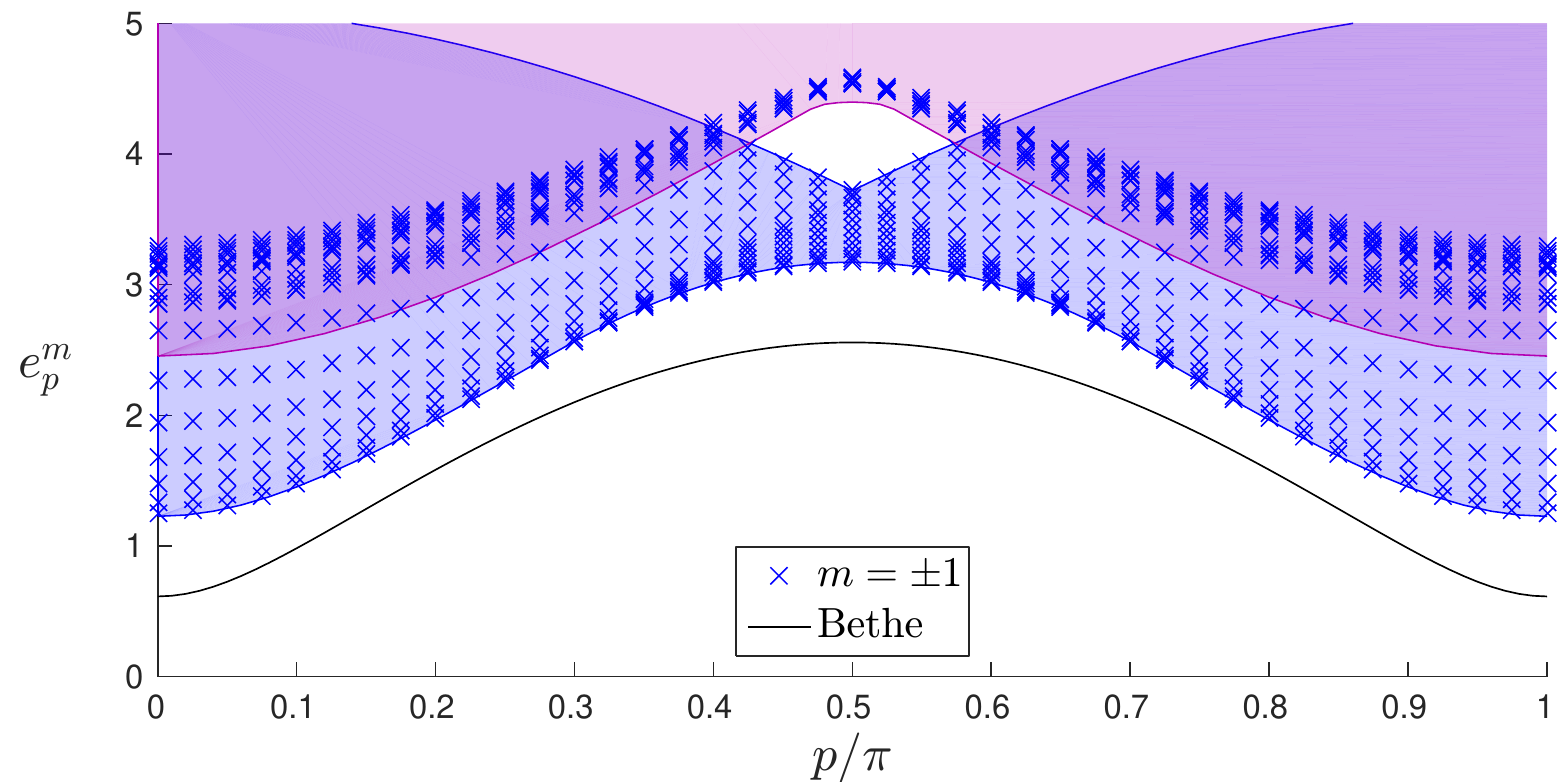}
 \includegraphics[width=0.495\linewidth,keepaspectratio=true]{\figpath/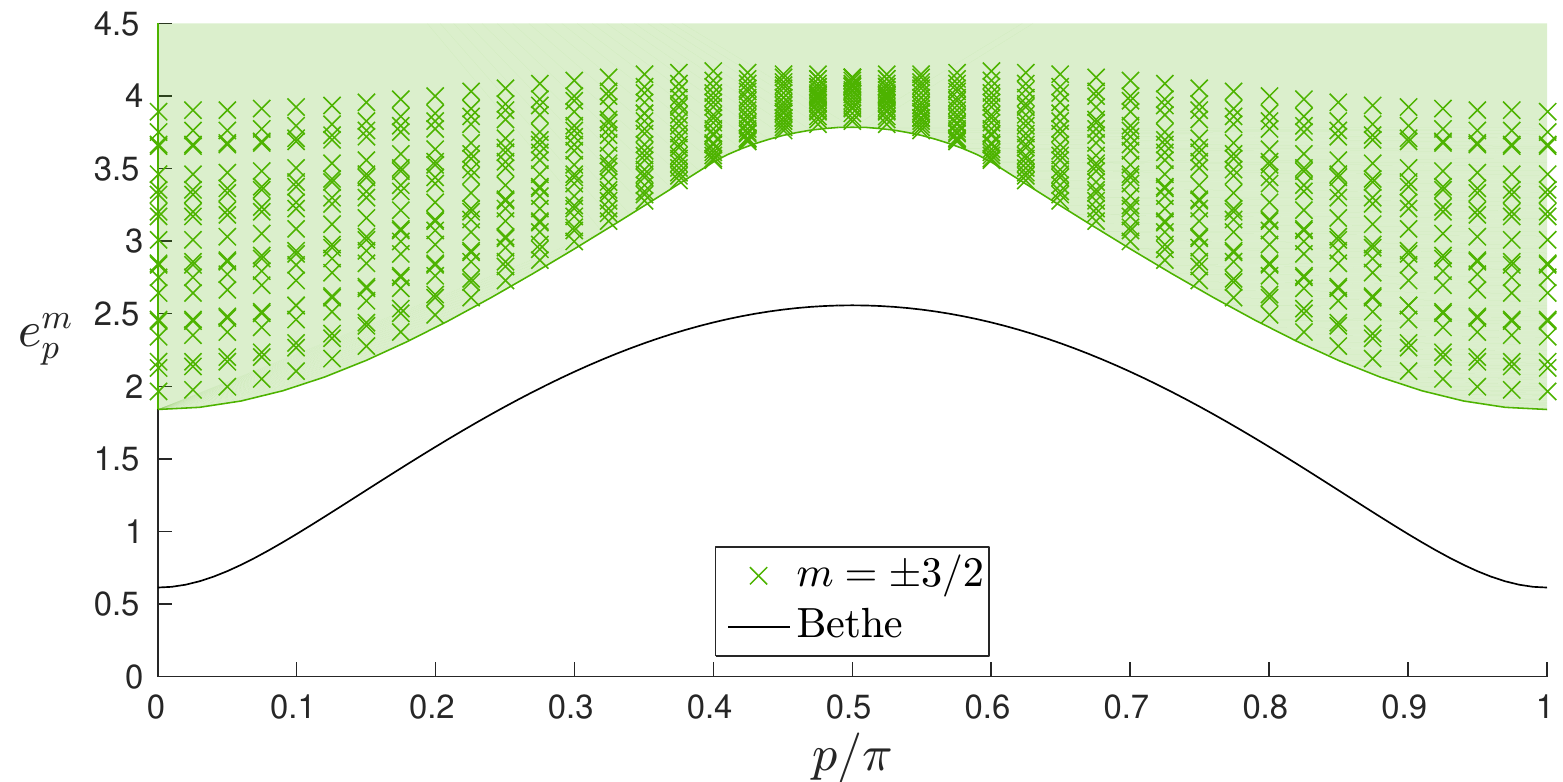}
 \includegraphics[width=0.495\linewidth,keepaspectratio=true]{\figpath/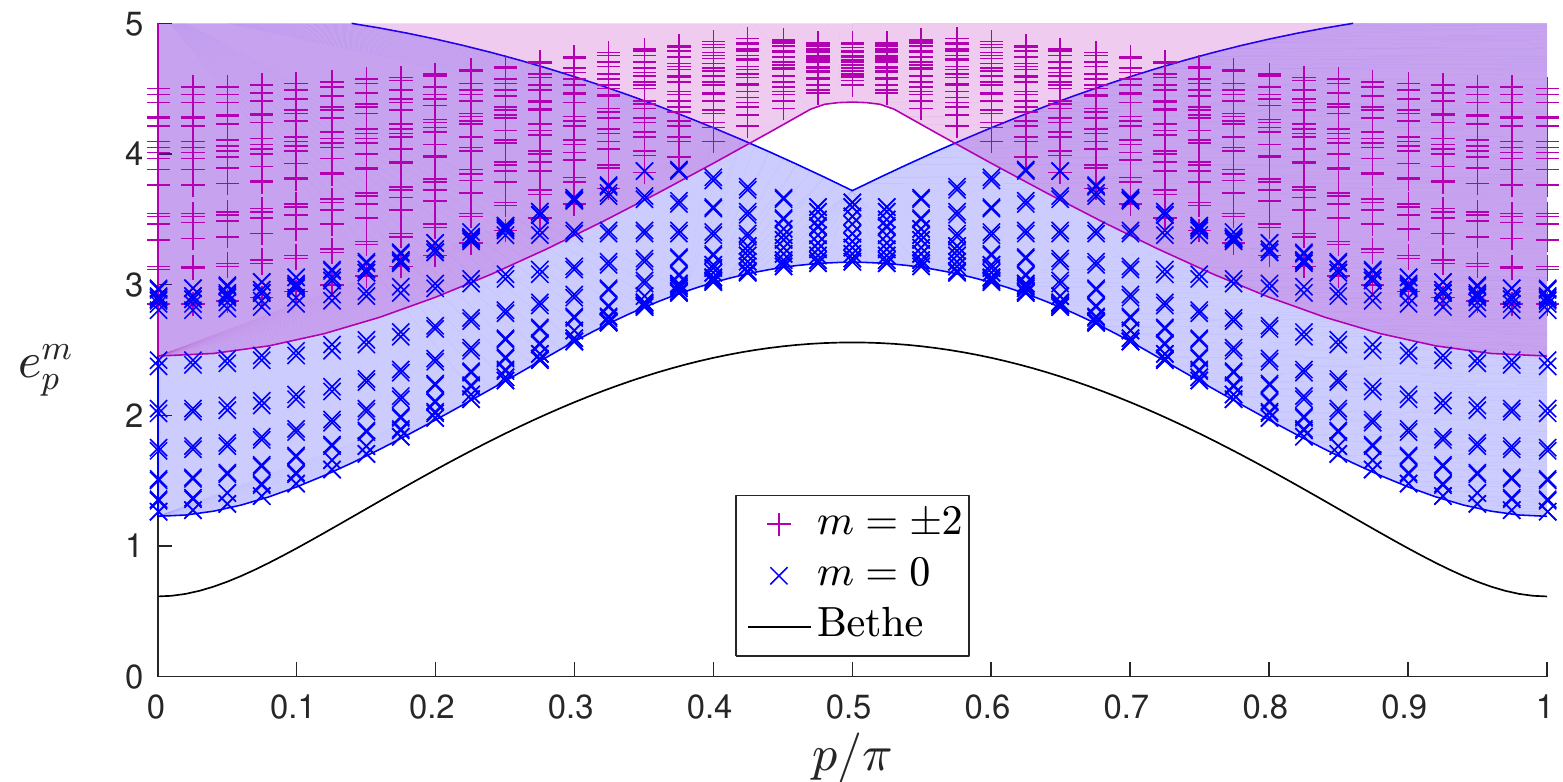}
 \caption{Variational low energy dispersion for the antiferromagnetic XXZ model \eqref{eq:XXZ_Ham} at $\Delta=3, h=0$ and bond dimension $D=70$. We show results for half integer magnetization (topologically nontrivial) excitations on the left, while results for integer magnetization (topologically trivial) excitations are shown on the right, together with exact results from Bethe ansatz,\cite{Takahashi,CloiseauxGaudin} where the black solid line denotes the exact elementary branch, and the blue, green and purple areas the exact 2, 3 and 4 particle scattering continua respectively. We show the first 20 lowest energies for each magnetization. While excitations in the multi particle continua get partially reproduced, the elementary spinon with $m=\pm1/2$ (lowest red branch in the top left panel) is reproduced to machine precision.}
 \label{fig:XXZ_excitations}
\end{figure*}

We consider the antiferromagnetic case $\Delta>0$. There the ground state in zero field has zero magnetization, and the elementary excitations are given by \textit{spinons}.\cite{CloiseauxGaudin, Takahashi} In contrast to simple spin flip excitations (magnons) -- which are integer spin excitations -- spinons have fractional spin $S=1/2$.\cite{Spinon}
In the following we will demonstrate that in the context of the symmetric ansatz \eqref{eq:msuc_excitation_ansatz}, spinon excitations must necessarily be of topologically nontrivial nature, i.e. they are domain wall like and cannot be created by a single (or few) local spin flips.

For a zero magnetization ($m_{0}=0$) suMPS ground state approximation, we require a unit cell size $N=2$. We use a shifted quantum number $\tilde{s}(\s)=s(\s)=\pm 1/2$, and the quantum number $\xqn$ of the excitation directly corresponds to the magnetization $m$ of the excitation.
\footnote{Here and in the following we loosely speak of an excitation's magnetization, which is to be understood as the magnetization difference to the ground state.}
Without loss of generality we assume integer quantum numbers on even bonds, and half-integer quantum numbers on odd bonds. Consider the contribution 
\begin{equation}
A(1)_{L}^{\s_{1}}B(2)^{[\xqn]\s_{2}}=A(1)_{L}^{\s_{1}}V(2)^{\s_{2}}x^{[\xqn]}.
\end{equation} 
For a topologically \textit{trivial} excitation, the next unit cell starts again with $\tilde{A}(1)_{R}=A(1)_{R}$. The outgoing quantum numbers of $V(2)$ and $A(2)$ however are the same, which are in turn also equal to the ingoing quantum numbers of $A(1)$, all of which are integers. This means that only integer values $\xqn$ are possible, such that both in- and outgoing quantum numbers of $x^{[\xqn]}$ are integers. The local (single-mode approximation like) nature of the ansatz thus leads to the well known fact that excitations generated by localized spin flips can only generate integer spin excitations, i.e. \textit{magnons} (where e.g. $\xqn=\pm1$ corresponds to single spin flips)

The only possibility for $\xqn$ to be half-integer in order to generate a \textit{spinon}, is for the unit cell $\tilde{\mathbbm{A}}$ to the right of $\mathbbm{B}$ to start with half-integer instead of integer quantum numbers. This can be achieved by using a translated unit cell for $\tilde{\mathbbm{A}}$,
i.e. $\tilde{A}(1)=A(2)$, $\tilde{A}(2)=A(1)$, or $\ket{\Psi(\tilde{\mathbbm{A}})}=T\ket{\Psi(\mathbbm{A})}$ with $T$ the (single site) translation operator. Due to translation invariance of \eqref{eq:XXZ_Ham}, $\ket{\Psi(\tilde{\mathbbm{A}})}$ is also a valid ground state approximation with the same energy as $\ket{\Psi(\mathbbm{A})}$. 

Indeed, in the gapped antiferromagnetic phase $\Delta>1$ the exact ground state of \eqref{eq:XXZ_Ham} is twofold degenerate and spontaneously breaks translation invariance with finite \textit{staggered} magnetization density $m_{s}=\frac{1}{L}\sum_{j}(-1)^{j}\braket{Z_{j}}\neq0$, and the above $\ket{\Psi(\tilde{\mathbbm{A}})}\neq \ket{\Psi(\mathbbm{A})}$ are good ground state approximations. 
For $-1<\Delta\leq1$ however the exact ground state is unique and the model is gapless. For finite $D$ the above suMPS ground state approximations however \textit{artificially} break translation invariance, such that still $\ket{\Psi(\tilde{\mathbbm{A}})}\neq \ket{\Psi(\mathbbm{A})}$ are ground state approximations with the same variational energy and $m_{s}\neq0$ and we can use them to build spinon excitations. This symmetry is restored in the limit $D\to\infty$, where $m_{s}\to 0$ and $|\braket{\Psi(\tilde{\mathbbm{A}})|\Psi(\mathbbm{A})}|\to1$.

Colloquially speaking, the variational ansatz for a spinon is therefore obtained by ``pulling'' the 2 site unit cell ground state approximation apart by one site and inserting a variational perturbation matrix $B$ carrying two half integer spins (one for the local physical $S=1/2$ spin carried by $V$, and one for the spinon excitation carried by $x$) and the resulting ansatz is topologically nontrivial. Due to the 2 site unit cell, the momentum of the spinon is also restricted to $0\leq p\leq\pi$, i.e. to half of the first Brillouin zone, which is also consistent with \Ref{Spinon, Takahashi}. 

Here, generating $\tilde{\mathbbm{A}}$ from translating $\mathbbm{A}$ also removes the above mentioned ambiguity of the excitation quantum number $\xqn$, arising from the fact that $\mathbbm{A}$ and $\tilde{\mathbbm{A}}$ can in principle have arbitrary differences in the global quantum number offset. For a graphical representation of the construction of topologically trivial (magnon) and nontrivial (spinon) excitations, see \Fig{fig:qn_excitations}. 

For the case $N=2$ and $m_{0}=0$ we calculate the variational excitation energy dispersion with integer and half integer magnetizations for the XXZ Antiferromagnet \eqref{eq:XXZ_Ham} at $\Delta=3, h=0$ for bond dimension $D=70$. The numerical results are shown in \Fig{fig:XXZ_excitations}, together with exact results from Bethe ansatz.\cite{Takahashi,CloiseauxGaudin} Excitations with half integer magnetizations are of topologically nontrivial nature, while topologically trivial excitations carry integer magnetization. The elementary excitations are given by spinons with magnetization $m=\pm1/2$, and the entire spectrum of excitations can be generated from compositions of multiple spinon states into scattering or bound states. 

As mentioned below \eqref{eq:excitation_ansatz}, the ansatz only captures elementary excitations and their bound states well. Consequently, the elementary spinon branch is accurate to machine precision, while excitations in multi particle continua are only partially reproduced. Nevertheless, the low energy boundaries of these continua are still surprisingly well reproduced, where accuracy however decreases quickly with higher particle number. An interesting advantage of the suMPS ansatz is that excitations very high up in the spectrum with high magnetizations (e.g. bound states) can now easily be targeted, whereas in the non-symmetric approach such states would be buried high up in a multitude of other states and targeting them would require more involved numerical procedures. This also enables a systematic estimation of unknown lower bounds of high up multi particle continua by targeting high magnetization excitations.

\subsection{Magnetic Field for Ground State from Excitations in the XXZ Model}
\label{sec:h_from_e}

\begin{figure}[t]
 \centering
 \includegraphics[width=\linewidth,keepaspectratio=true]{\figpath/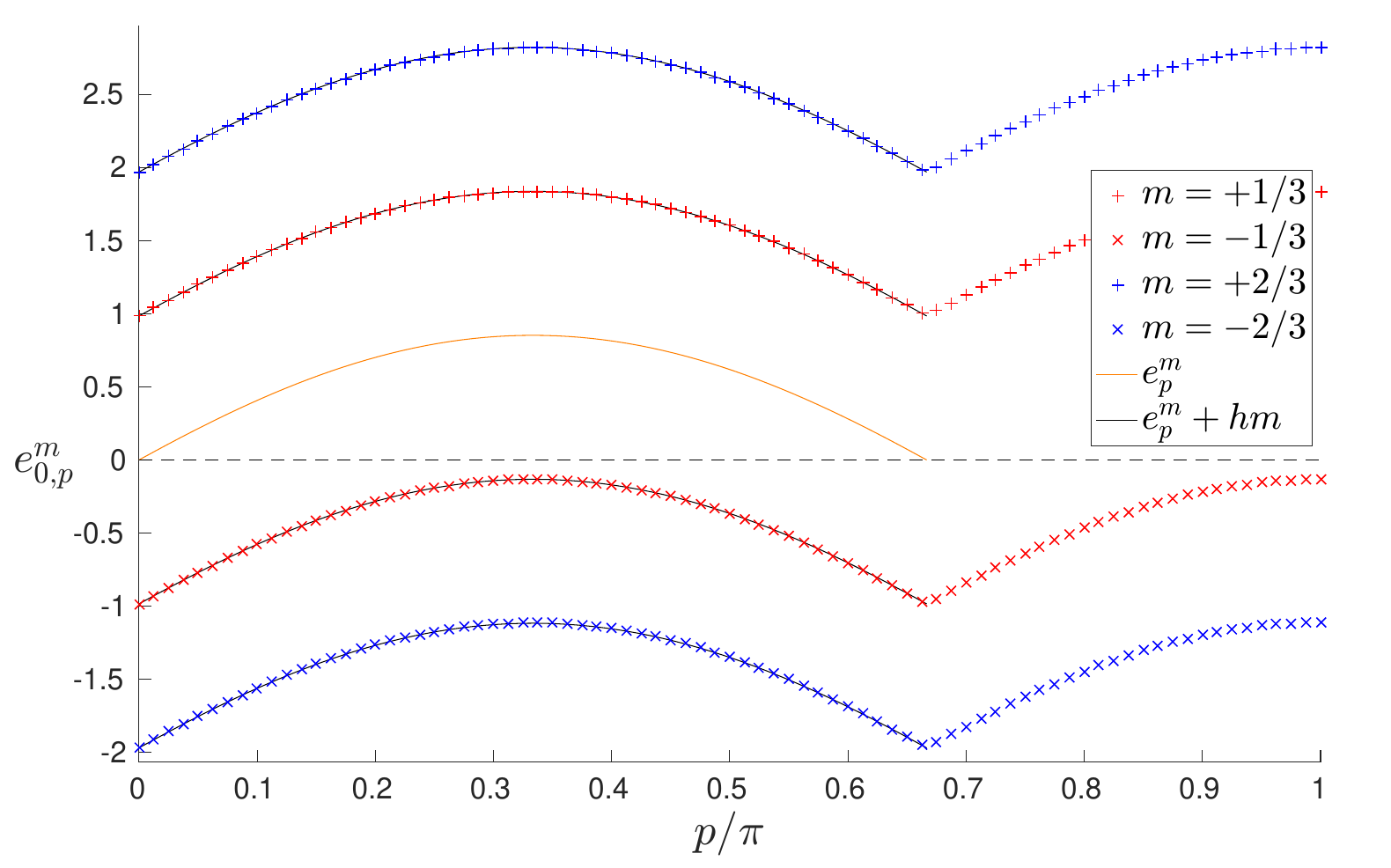}
 \caption{Dispersion relations $e^{m}_{0,p}$ of elementary excitations with fractional magnetizations $m=\pm 1/3$ (red symbols) and $m=\pm2/3$ (blue symbols) on top of a $m_{0}=1/6$ lowest energy state $\ket{\Psi_{0}}$ with respect to $H_{0}$. The energies are shifted by $\Delta e=hm$ with respect to the excitation energy dispersion $e^{m}_{p}$ of $H_{h}$, for which $\ket{\Psi_{0}}$ is the overall ground state. The solid lines show the exact dispersion $e^{m}_{p}$ obtained from Bethe ansatz\cite{Takahashi,CloiseauxGaudin} for reference.}
 \label{fig:XXZ_Delta3_m16}
\end{figure}

Apart from directly targeting and identifying excitations with certain quantum numbers, we can also use suMPS excitations to calculate the magnetic field $h$ required for the ground state of \eqref{eq:XXZ_Ham} to have a certain magnetization density $m_{0}$ (or total magnetization $M=Lm_{0}$). This is possible, as the magnetic field term $H_{M}=\sum_{j}Z_{j}$ commutes with the rest of the Hamiltonian $H_{0}$, and changing $h$ just changes the eigenenergies, but not the eigenstates of \eqref{eq:XXZ_Ham}. 

More specifically, we write $H_{\rm XXZ}=H_{h}={H}_{0}-hH_{M}$ and 
assume $\ket{\Psi_{0}^{M}}$ to be the ground state of $H_{h}$ with ground state energy $E^{(0)}_{h}$ and (total) magnetization $M$ for a suitable (unknown) value of $h$.
For a low lying excitation $\ket{\Phi^{m}_{\mom}}$ of $H_{h}$ with magnetization $m$  and momentum $\mom$ we have
\begin{align}
 (H_{h}-E^{(0)}_{h})\ket{\Phi^{m}_{\mom}}&=e^{m}_{\mom}\ket{\Phi^{m}_{\mom}}\\
 (H_{0}-E_{0})\ket{\Phi^{m}_{\mom}}&=\underbrace{(e^{m}_{\mom}+hm)}_{e^{m}_{0,\mom}}\ket{\Phi^{m}_{\mom}}\label{eq:H0_EV}
\end{align}
with 
$E_{0}=E^{(0)}_{h}+hM$ the corresponding eigenenergy of $H_{0}$. 
Note that while $e^{m}_{\mom}\geq0$ here the excitation energy $e^{m}_{0,\mom}$ with respect to $H_{0}$ need not be positive. In fact, as a function of $\mom$ we obtain the true excitation energy dispersion $e^{m}_{\mom}$ of $H_{h}$ shifted by a constant energy offset $hm$.

Let us now obtain the overall ground state $\ket{\Psi_{0}^{M}}$ of $H_{h}$ as the lowest energy state with magnetization $M$ and energy $E_{0}$ from $H_{0}$. In order to infer $h$ we then construct a variational excitation $\ket{\Phi^{m}_{\mom}}$ on top of $\ket{\Psi_{0}^{M}}$. Its variational energy $e^{m}_{0,\mom}$ with respect to $H_{0}$ is then given by \eqref{eq:H0_EV}, from which we can calculate $h$ if we know $e^{m}_{\mom}$. 

From symmetry arguments we however know that $e^{m}_{\mom}=e^{-m}_{\mom}$, such that we can additionally construct $\ket{\Phi^{-m}_{\mom}}$ and use both variational excitation energies to obtain
\begin{equation}
e^{m}_{\mom}=(e^{m}_{0,\mom} + e^{-m}_{0,\mom})/2. 
\label{eq:em_from_pm}
\end{equation} 

\begin{table}[t]
  \begin{tabular}{r|r|r}
\multicolumn{1}{c|}{$m$} &\multicolumn{1}{c|}{$\Delta=0.5$}&\multicolumn{1}{c}{$\Delta=3$}\\
  \hline
  1/3 &   0.2769651864 &  0.9850346057\\
 -1/3 &  -0.2758610821 & -0.9835478736\\
  2/3 &   0.5535936671 &  1.9698464387\\
 -2/3 &  -0.5520823216 & -1.9673553352
 \end{tabular} 
 \caption{Variational energies $e^{m}_{0,\mom}$ with respect to $H_{0}$ for elementary excitations with fractional magnetizations $m=\pm1/3, \pm2/3$ and bond dimension $D=200$ for momentum $p=0$ on top of a $m_{0}=1/6$ lowest energy state of $H_{0}$.}
\label{tab:eq_m13}
\end{table} 

\begin{table}[t]
  \begin{tabular}{r|r|r}
\multicolumn{1}{c|}{$m$} &\multicolumn{1}{c|}{$\Delta=0.5$}&\multicolumn{1}{c}{$\Delta=3$}\\
  \hline
  1/3   &   0.8292394026 &  2.9528737189\\
  2/3   &   0.8292569915 &  2.9529013304\\
  \hline
  exact &   0.8291610777 &  2.9529130736\\
 \end{tabular} 
 \caption{Calculated magnetic fields $h$ necessary for the ground state of \eqref{eq:XXZ_Ham} to have magnetization density $m_{0}=1/6$. Shown are the values calculated from the data in \Tab{tab:eq_m13} and Eqns.~\eqref{eq:em_from_pm} and \eqref{eq:h_from_em} for $m=1/3,2/3$, together with exact values from Bethe Ansatz.\cite{Takahashi,CloiseauxGaudin}}
\label{tab:h_m13}
\end{table} 

We demonstrate this method to calculate the magnetic field $h$ required for a ground state with magnetization density $m_{0}=1/6$ in the antiferromagnetic cases $\Delta=0.5$ and $\Delta=3$.
We start by obtaining the lowest energy state of $H_{0}$ with $m_{0}=1/6$, which requires a $N=3$ site unit cell suMPS representation. Due to a finite bond dimension representation the obtained ground state approximation for $\Delta=0.5$ also (artificially) breaks translation invariance, and for both values of $\Delta$ there are three different and equally good ground state approximations which are related by single or two site translations. From these states we can construct several topologically nontrivial elementary excitations by combining the different translated unit cells left and right of $B$ in an excitation ansatz with (fractional) quantum numbers $\xqn=\pm 1/3, \pm 2/3$. There are 3 possibilities for each quantum number, totaling in 12 possible states.
\footnote{In the general case of any translation invariant Hamiltonian with $N$ different $N$ site unit cell ground state approximations with equal variational ground state energy, we can in principle construct $2N(N-1)$ different (but perhaps equivalent) topologically nontrivial excitations with fractional quantum numbers:
There are $N(N-1)$ possible ordered ways to combine $N$ different unit cells, and for each combination there is one excitation with a positive and one with a negative fractional quantum number. See also Ref. \onlinecite{Fractional_Spinon_Okunishi}.}
See \App{sec:m13_excitations} for more details on obtaining excitations energies for well defined (fractional) magnetizations.

Without loss of generality we choose momentum $\mom=0$ and calculate variational excitation energies $e^{m}_{0,\mom}$ with respect to $H_{0}$ with $D=200$ and fractional magnetizations $m=\pm1/3,\pm2/3$ for $\Delta=0.5,3$. 
\footnote{Instead of domain wall excitations with fractional magnetization we could have also used topologically trivial magnon excitations with integer magnetization. These excitations are in general however scattering states of two or three elementary excitations, and the accuracy of their variational energy will be worse than the accuracy for elementary excitations.}
The numerical results are given in \Tab{tab:eq_m13}. From these values we further obtain the true excitation energies $e^{m}_{\mom}$ with respect to $H_{h}$ from \eqref{eq:em_from_pm}. These energies are known to be exactly zero\cite{Takahashi,CloiseauxGaudin} and we obtain values of the order $\O(10^{-4})$, due to finite bond dimension.
Finally, we calculate $h$ from
\begin{equation}
 h = (e^{m}_{0,\mom}-e^{m}_{\mom})/m
 \label{eq:h_from_em}
\end{equation} 
The numerically obtained values for $h$ are given in \Tab{tab:h_m13}. In comparison to the exact values from Bethe Ansatz\cite{Takahashi,CloiseauxGaudin} the errors are of order $\O(10^{-5})$ and thus quite low a for moderate bond dimension of $D=200$. The (shifted) variational energy dispersions are shown in \Fig{fig:XXZ_Delta3_m16}.

This method is not restricted to determining necessary magnetic fields $h$ for a certain ground state magnetization (density), but is generally applicable to all Hamiltonians which contain one (or more) generators of their global symmetries as a parameterized term. It is especially useful for models that are not exactly solvable, where otherwise one would have to perform a large number of ground state calculations in a grid search with small variations of $h$, while here $h$ is calculated \textit{directly} from a single ground state calculation followed by two (or few) excited state calculations.

\subsection{Spinons, Holons and Electrons in the Fermi Hubbard Model}
\label{sec:HUB_results}

\begin{figure*}[t]
 \centering
 \includegraphics[width=0.49\linewidth,keepaspectratio=true]{\figpath/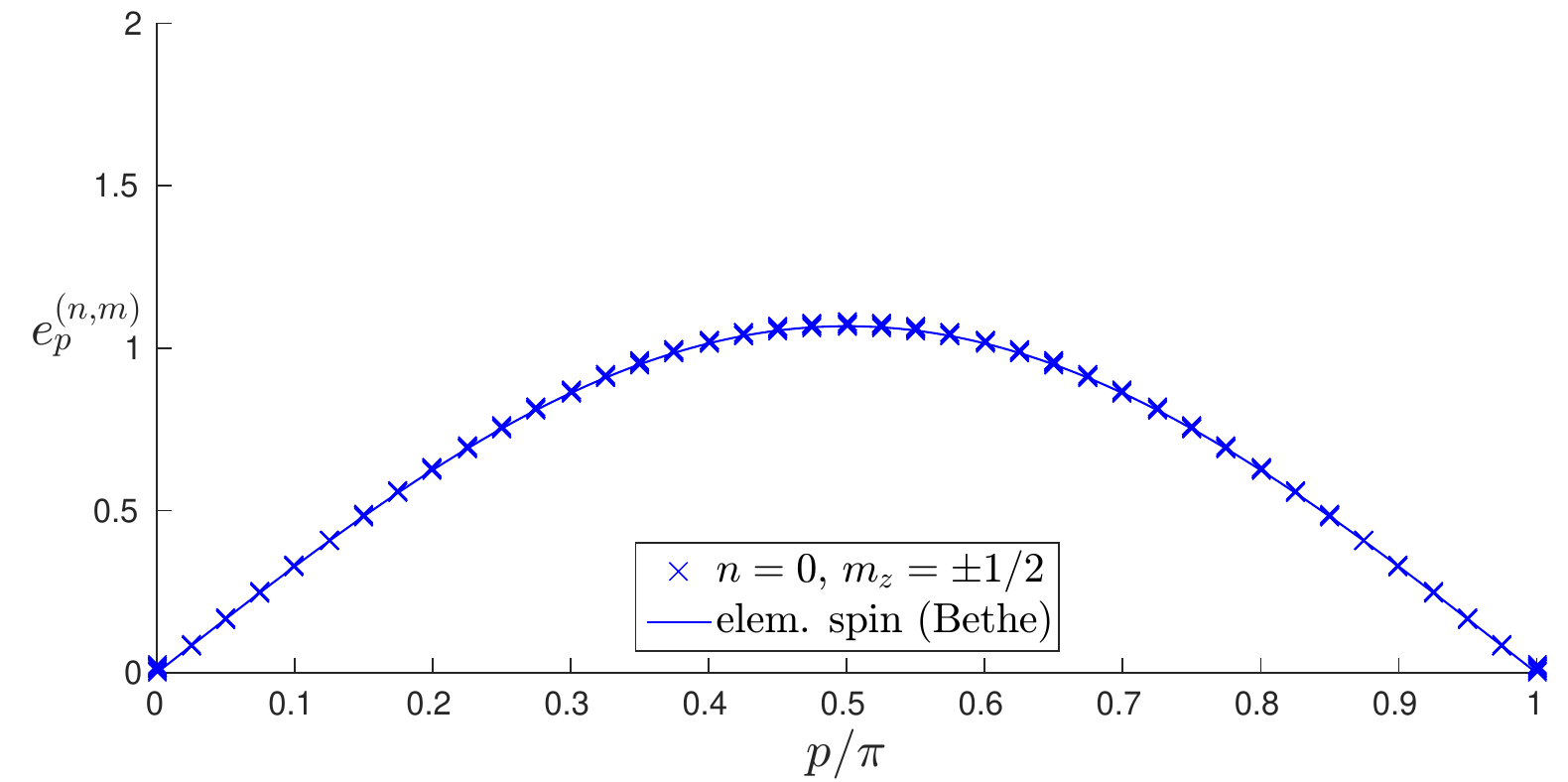}
 \includegraphics[width=0.49\linewidth,keepaspectratio=true]{\figpath/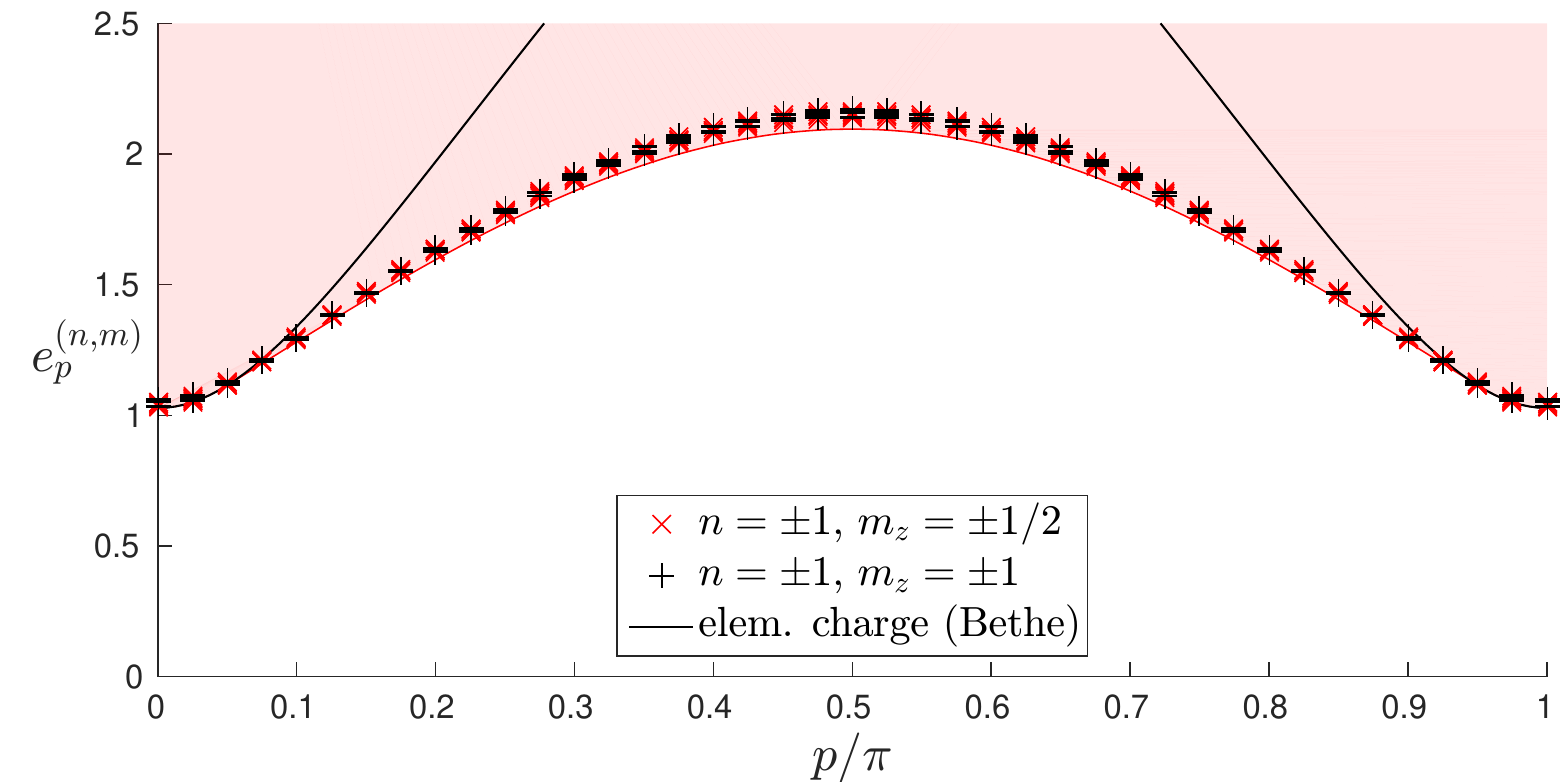}
 \includegraphics[width=0.49\linewidth,keepaspectratio=true]{\figpath/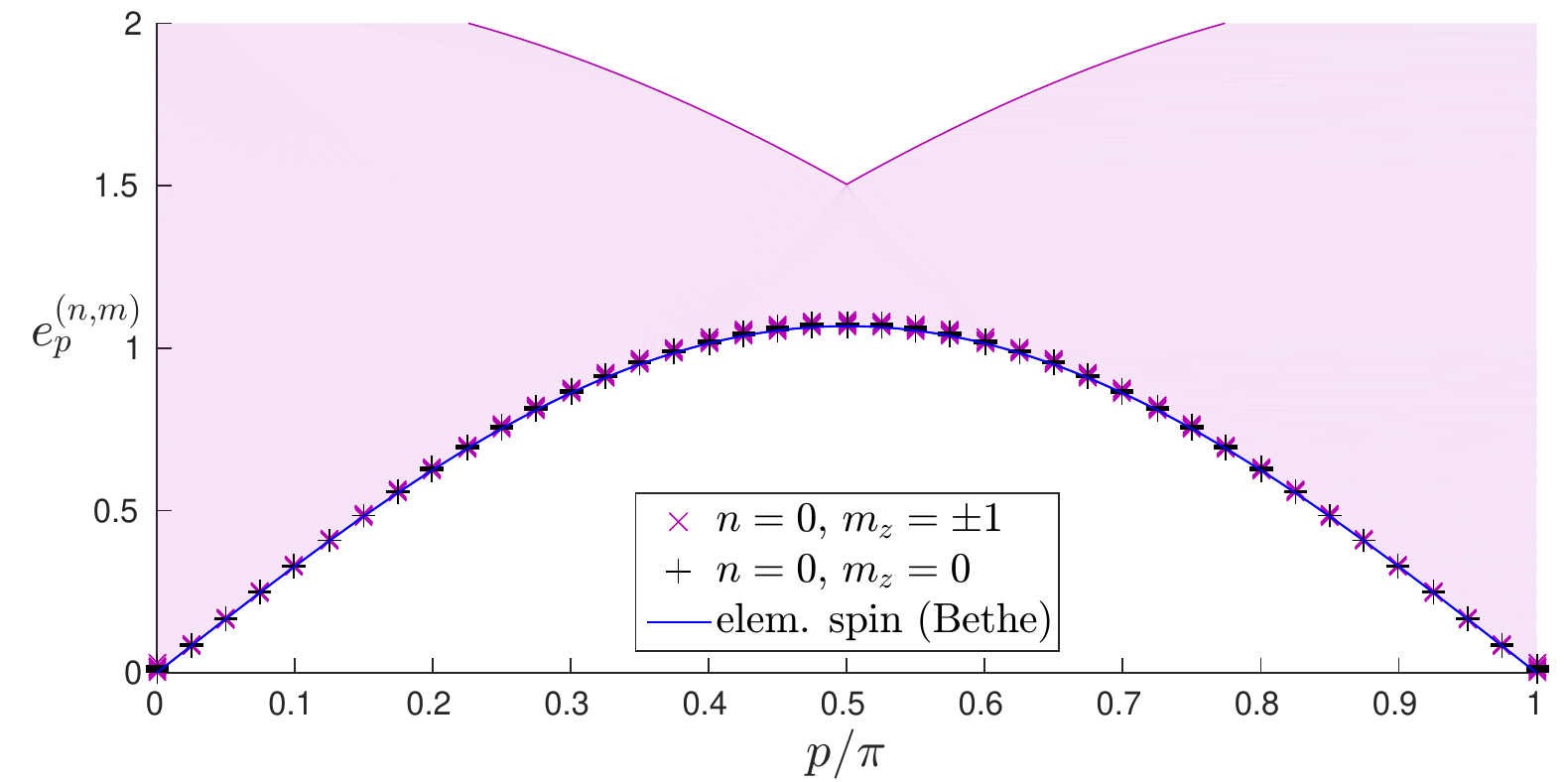}
 \includegraphics[width=0.49\linewidth,keepaspectratio=true]{\figpath/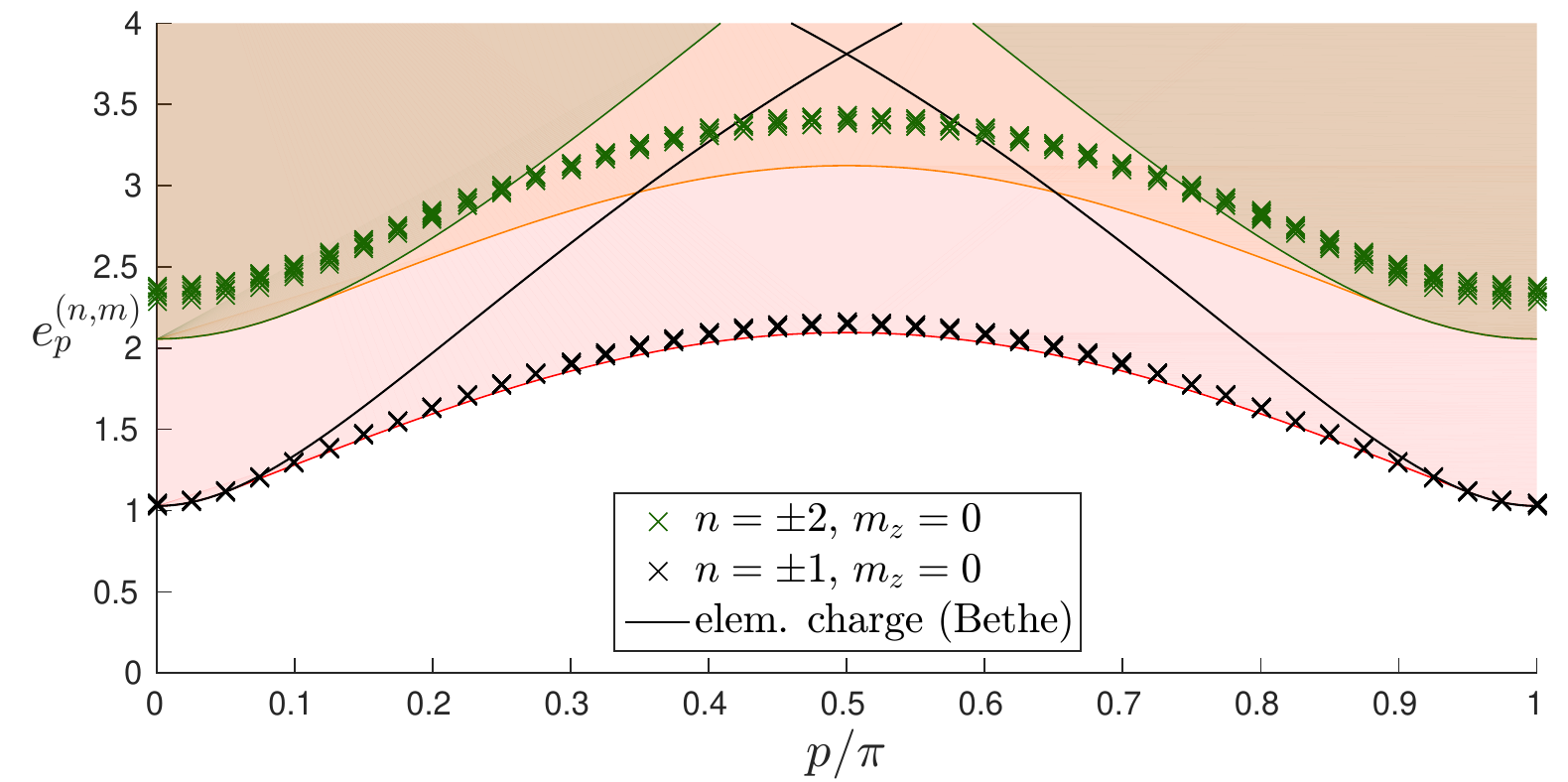}
 \caption{Variational low energy dispersion for the Fermi Hubbard model \eqref{eq:HUB_Ham} in the integrable case $V=0$ at $U=5$ and half filling ($n_{0}=1$, $m_{0}=0$), for bond dimension $D=600$. We show results for pure spin excitations on the left, while results for charge and spin-charge excitations are shown on the right. We show the first 8 lowest obtained variational energies for each quantum number. We also show the exact elementary branches and multi-particle continua from Bethe ansatz,\cite{Essler} where on the left the purple area is the continuum of spin-spin excitations, while on the right the green, red and orange areas correspond to charge-charge, spin-charge and spin-charge-charge excitation continua respectively. }
 \label{fig:HUB_excitations}
\end{figure*}

In the following two sections we consider as a second example the low energy spectrum of the (extended) Fermi Hubbard model
\begin{equation}
\begin{split}
 H_{\rm HUB}=&-t\sum_{\sigma,j}c_{\sigma,j}c^{\dagger}_{\sigma,j+1} - c^{\dagger}_{\sigma,j}c_{\sigma,j+1}\\
 &+ U\sum_{j}\left( n_{\uparrow,j}-\frac{1}{2} \right)\left( n_{\downarrow,j}-\frac{1}{2} \right)\\
 &+ V\sum_{j}\left( n_{j}-1 \right)\left( n_{j+1}-1 \right)
 -\mu\sum_{j}n_{j},
 \end{split}
 \label{eq:HUB_Ham}
\end{equation}
where $c_{\sigma,j}$, $c^{\dagger}_{\sigma,j}$ are creation and annihilation operators of electrons of spin $\sigma$ on site $j$, $n_{\sigma,j}=c^{\dagger}_{\sigma,j}c_{\sigma,j}$ and $n_{j}=n_{\uparrow,j} + n_{\downarrow,j}$ are the particle number operators. Here, $t$ is the hopping amplitude, $U$ and $V$ are the on site and nearest neighbor interactions respectively and $\mu$ is the chemical potential (we do not consider an external magnetic field).

Due to the phenomenon of spin charge separation,\cite{Tomonaga50,Luttinger63,Haldane81} the elementary excitations are fractionalized quasi-particles of either spin or charge alone, which cannot be constructed from the bare electrons, as those carry \textit{both} spin and charge. Rather, electrons can in turn be interpreted as bound states of these elementary spin and charge excitations, known as \textit{spinons} and \textit{holons} respectively. Consequently, we use quantum number representation $(n,m)$ of the local physical electronic states, with $n$ the particle number and $m$ the magnetization. Spinons and holons  carry quantum numbers $q_{s}=(0,\pm1/2)$ and $q_{c}=(\pm1,0)$ and excitations with electronic quantum numbers $q_{e}=(\pm1,\pm1/2)$ are therefore combinations of holons and spinons.

We first focus on the integrable case $V=0$, where the ground state energy and elementary excitations are known exactly from Bethe ansatz,\cite{Lieb_Wu68,Essler} at half filling $(n_{0},m_{0})=(1,0)$, which requires a two site unit cell for a suMPS ground state approximation. Much similar to the case of the elementary spinon in the XXZ model in \Sec{sec:XXZ_results}, topologically trivial excitations can only carry electronic quantum numbers, which are unsuitable for creating elementary spinons or holons, and the elementary excitations are thus necessarily topologically nontrivial also in this case.

For the suMPS ground state approximation at half filling we use the corresponding shifted quantum numbers given in \Tab{tab:qn_hub}. Even though the exact ground state is unique for all values of $U>0$, the finite bond dimension suMPS ground state approximation again artificially breaks translation invariance, and the translated ground state $\ket{\Psi(\tilde{\mathbbm{A}})}=T\ket{\Psi(\mathbbm{A})}\neq\ket{\Psi(\mathbbm{A})}$ is an equally good ground state approximation. From these we can now construct topologically nontrivial excitations which allow for quantum numbers $q_{c}$ and $q_{s}$ and we can generate elementary spinons and holons that way. Conversely, topologically trivial excitations carry quantum numbers that are even combinations of $q_{s}$ and $q_{c}$, for example spin-spin, charge-charge, or spin-charge excitations.

For the case $N=2$ and half filling we calculate variational excitation energy dispersions for $U=5$ and bond dimension $D=600$ for several different quantum numbers. There, the elementary charge excitations are gapped, while elementary spin excitations are gapless.\cite{Lieb_Wu68,Essler} The numerical results are shown in \Fig{fig:HUB_excitations}, together with exact results from Bethe ansatz, where we show pure spin excitations in the left plot and excitations with nonzero charge quantum numbers in the right plot. We find that the elementary spinon is reproduced up to an excellent accuracy of $\O(10^{-6})$ by the lowest variational excitation branch with $m=\pm1/2$. The higher up branches for integer and half integer magnetizations lie within the continuum of spin-spin scattering states which are not well captured by our ansatz. Nevertheless the variational energies reproduce the low end of this continuum surprisingly well.

Due to the elementary spinon being gapless, the exact elementary holon branch lies completely within the continua of scattering states of one charge and arbitrarily many spin excitations. Out of these, e.g. the charge-spin-spin continuum (red area in \Fig{fig:HUB_excitations}) also contains excitations with quantum number $\xqn=(\pm1,0)$ (charge + spin singlet or triplet with $m_{z}=0$) equal to $q_{c}$. The variational excitation ansatz for this $\xqn$ tries to reproduce exactly these excitations, as due to the smaller bandwidth of the spinon branch they are at lower energies than the elementary charge excitations (shown as black solid lines in \Fig{fig:HUB_excitations} on the right), except at $\mom=0,\pi$, where the lower bound of the continuum is exactly the elementary charge branch. Around these momenta, the variational ansatz with $\xqn=(\pm1,0)$ indeed yields the lowest energies and reproduces the elementary holon up to an accuracy of $\O(10^{-5})$. Away from $\mom=0,\pi$ the same ansatz tries to reproduce a three particle scattering state, and energies for $\xqn=(\pm1,\pm1/2)$ -- which try to reproduce two particle spin-charge scattering states -- yield in fact slightly lower energies. 
A similar effect can be observed for $\xqn=(\pm2,0)$ excitations, where there is a noticeable bend in the dispersion around $\mom\approx0.15$. There the ansatz tries to reproduce a spin-spin-charge-charge instead of a charge-charge scattering state.
\footnote{The fact that the spinon bandwidth is always smaller than the holon bandwidth is a special consequence of the Bethe solution for the integrable Hubbard model, but seems not to be a generic feature of electronic systems. In fact, away from integrability, such systems can even show much more diverse elementary excitations, with even more complex interactions (see e.g. \Sec{sec:results_EHUB}). }

Also, the exact elementary charge branch spans the entire Brillouin zone $\mom\in[0,2\pi)$,\cite{Essler} while the momentum of a two-site ansatz is necessarily restricted to half of the Brillouin zone. However, it turns out that the two-site unit cell is required only by the spin quantum numbers, while half filling for charge quantum numbers alone could be achieved with a single site unit cell ansatz. Consequently, charge excitations are reproduced twice by this ansatz, with a relative shift of $\mom=\pi$ over the entire Brillouin zone. For that reason we also draw the exact elementary charge branch from Bethe ansatz twice with the corresponding momentum shift in \Fig{fig:HUB_excitations}.

A possible way to remedy this fact is to realize that the variational ground state -- even though not translation invariant under pure translations $T$ -- is invariant under a translation followed by a spin flip, i.e. under application of $TF_{S}$, where $T$ is the translation and $F_{S}$ is the spin flip operator. The entire Brillouin zone for charge excitations could therefore be recovered by using a restricted ansatz $B(2) = \rme^{\rmi\mom}F_{S}B(1)$,
which indeed yields an eigenstate of $TF_{S}$ with eigenvalue $\rme^{-\rmi\mom}$ where $\mom\in[0,2\pi)$ can be interpreted as quasi-momentum covering the full Brillouin zone.

Additional ways to only target the elementary charge branch within the charge-spin scattering continuum would be to either minimize the energy \textit{variance} of the variational ansatz (see \onlinecite{Laurens_Scattering_2}, especially Appendix 4), instead of the energy itself, or to distinguish excited states by higher conservation laws,\cite{HubLadderOps1,HubLadderOps2} which could be introduced as artificial penalty terms into the Hamiltonian. The latter option however requires analytical knowledge about these higher conserved quantities and is unsuited for non-integrable models. We leave all these avenues to be explored in future studies. 

Overall it is demonstrated that the new symmetric suMPS ansatz allows for an efficient separation of excitation sectors with different quantum numbers, which was possible in the original proposal in \Ref{MPS_Excitations_variational} only a posteriori and with great effort. For example, charge excitations only start appearing above the $U$ dependent charge gap above a continuum of pure spin excitations. In the non-symmetric original ansatz these excitations are next to impossible to single out or target, especially if the value of the charge gap is unknown. 

In fact, even despite the elementary charge branch lying completely within multi-particle continua, the charge gap can still be calculated with the new symmetric ansatz to excellent precision of the same order as the ground state.

\subsection{Spin to Charge Density Wave Phase Transition in the Half Filled Extended Fermi Hubbard Model}
\label{sec:results_EHUB}

\begin{figure*}[p]
 \centering
 \includegraphics[width=0.495\linewidth,keepaspectratio=true]{\figpath/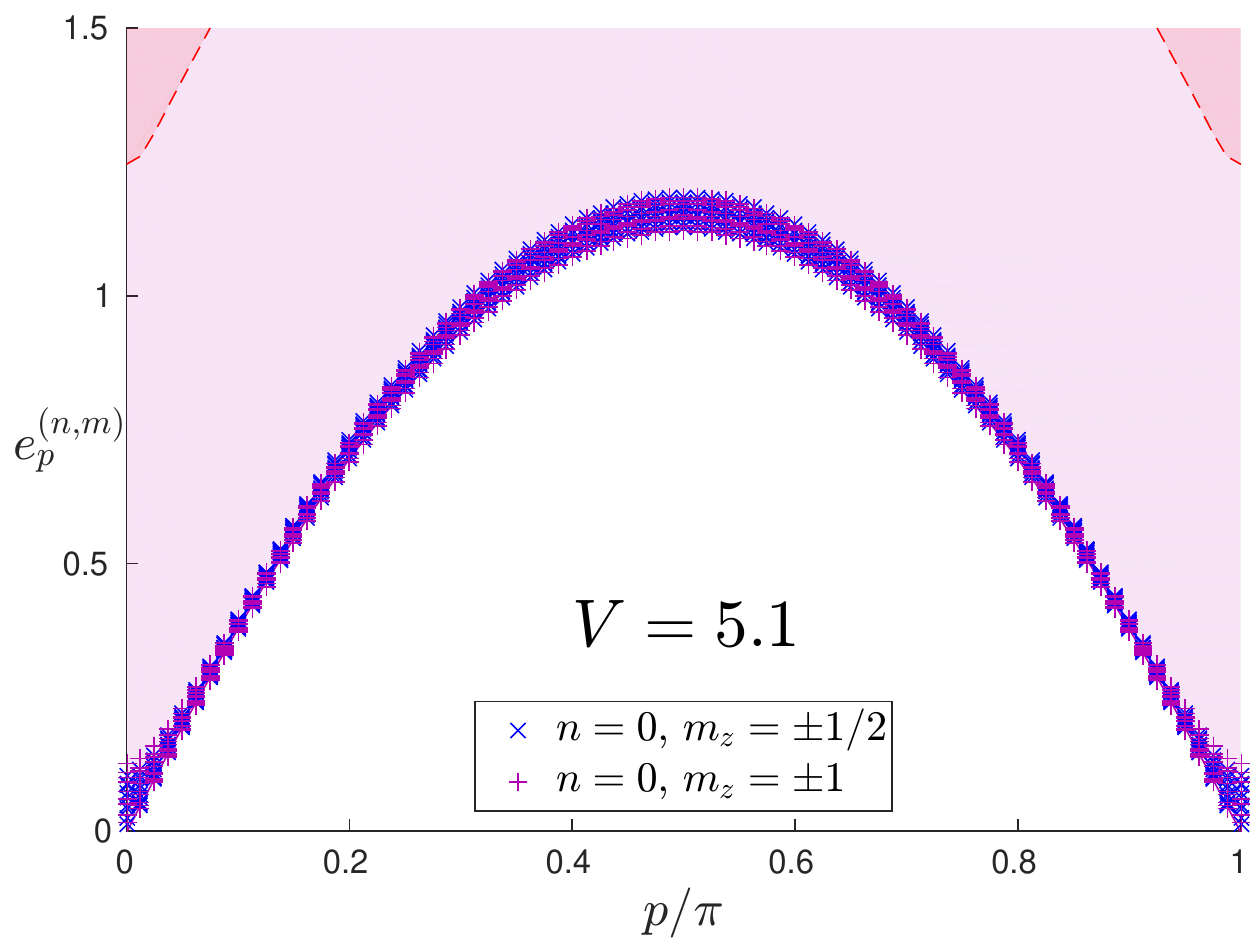}
 \includegraphics[width=0.495\linewidth,keepaspectratio=true]{\figpath/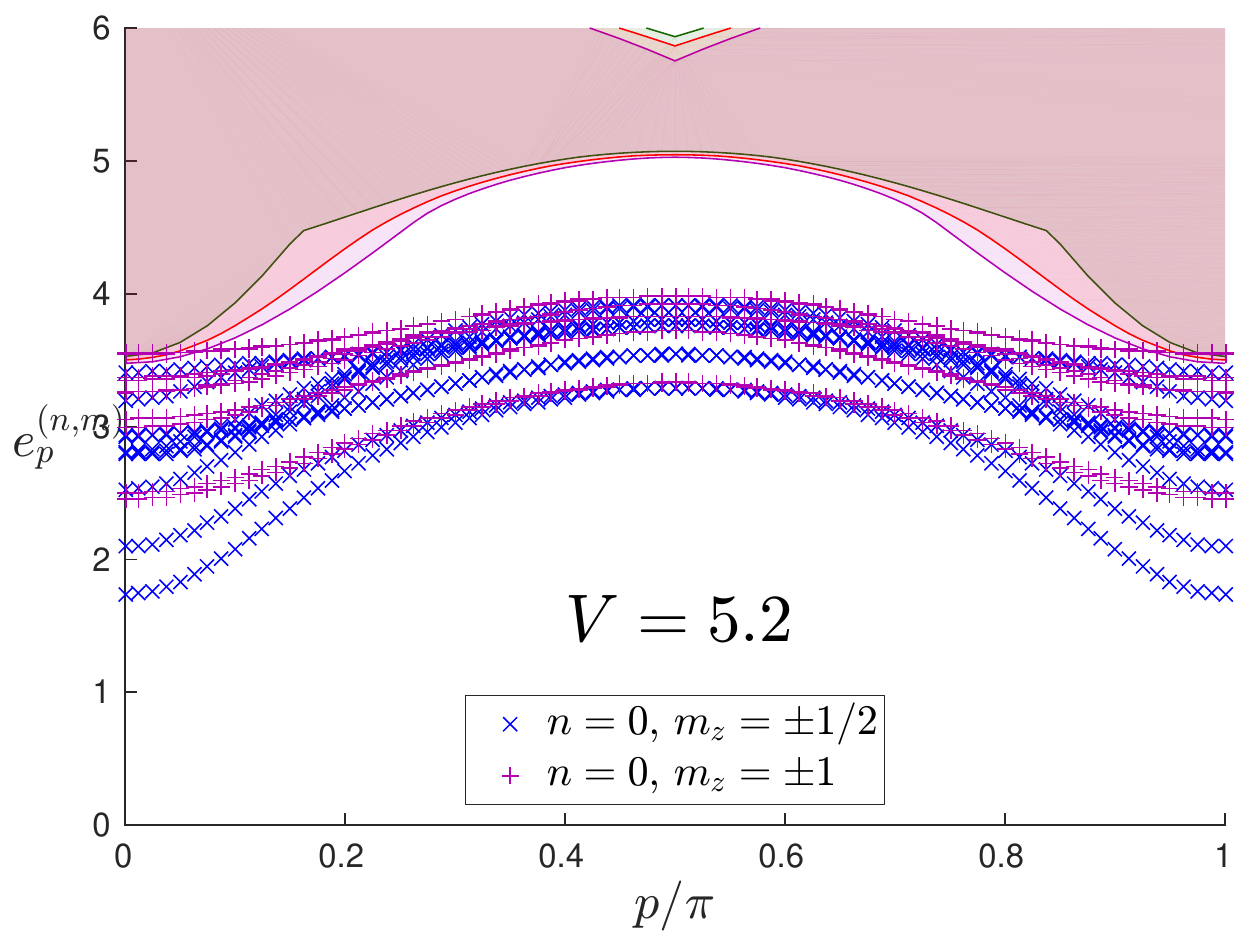}
 \includegraphics[width=0.495\linewidth,keepaspectratio=true]{\figpath/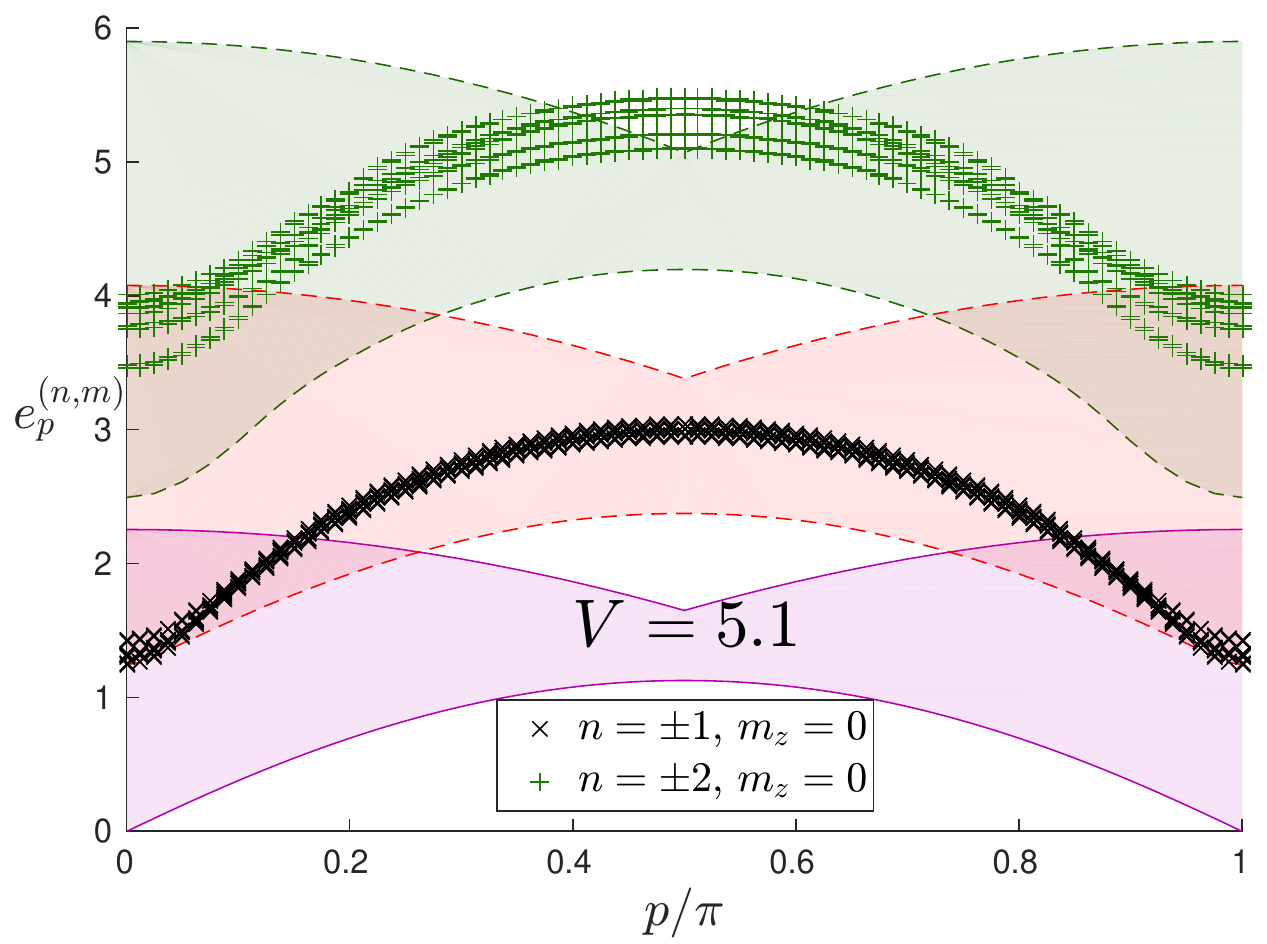}
 \includegraphics[width=0.495\linewidth,keepaspectratio=true]{\figpath/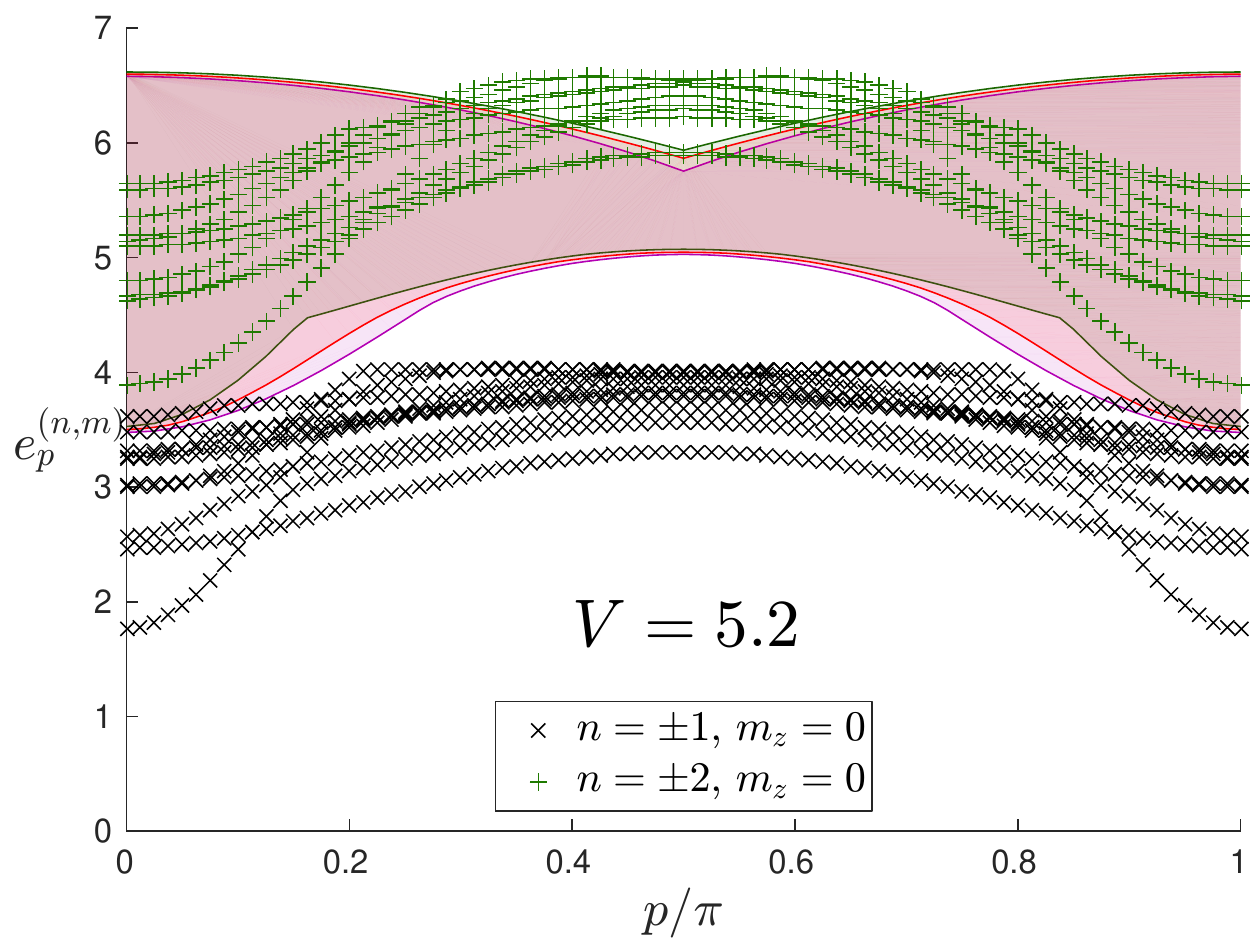}
 \includegraphics[width=0.495\linewidth,keepaspectratio=true]{\figpath/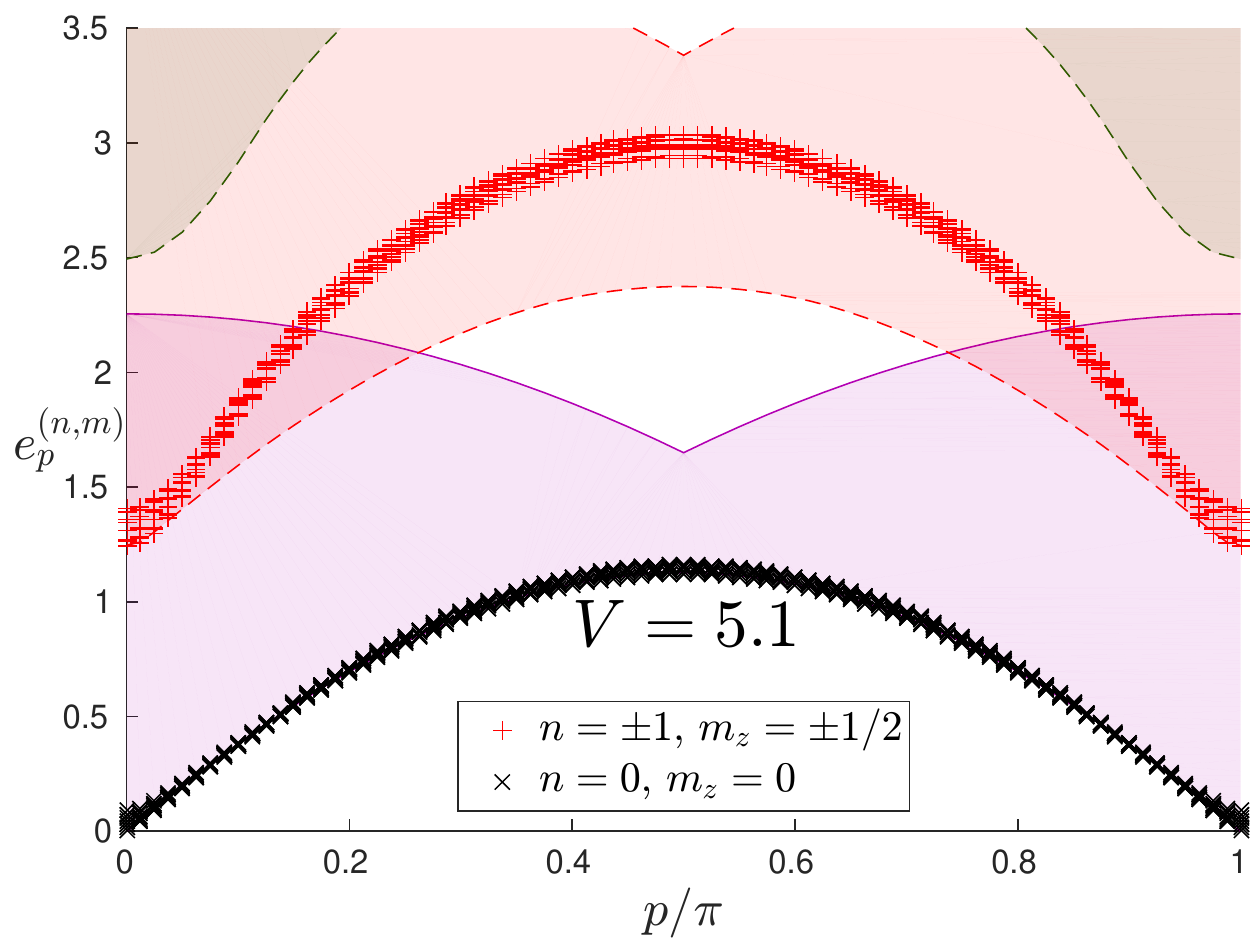}
 \includegraphics[width=0.495\linewidth,keepaspectratio=true]{\figpath/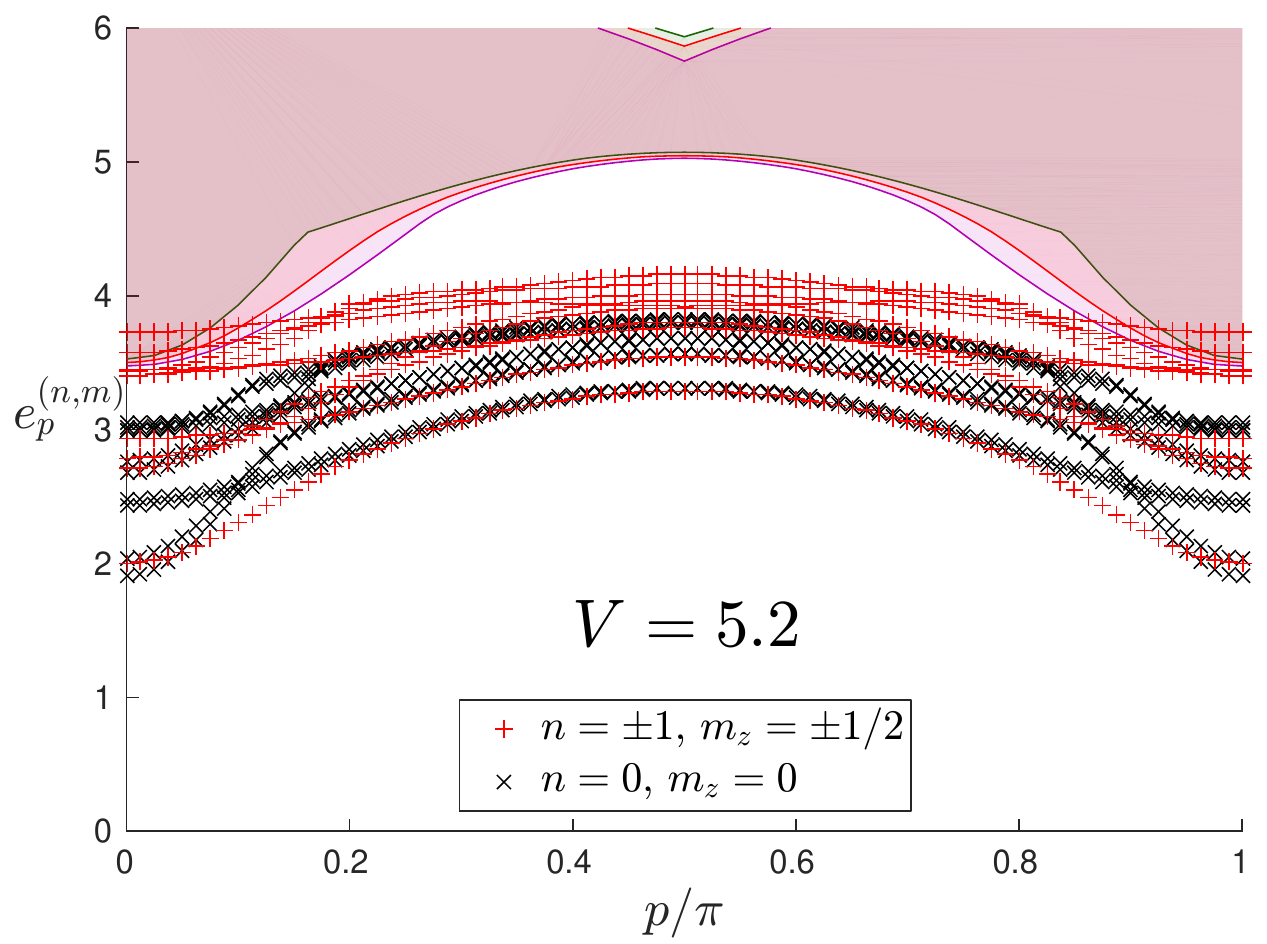}
 \caption{Variational low energy dispersions for various quantum numbers of the extended Fermi Hubbard model \eqref{eq:HUB_Ham} in the non integrable case $V=5.1$ (left) and $V=5.2$ (right) at $U=10$ on top of a half filled ground state ($n_{0}=1$, $m_{0}=0$). We show the first 10 lowest energies for each quantum number, represented by colored symbols. In addition we show spin-spin, charge-charge and spin-charge scattering continua constructed from the variational elementary spin and charge excitations as purple, green and red areas respectively. The spectrum on the left is in the SDW phase $V<V_{c}\approx 5.13$, while the spectrum on the right is in the CDW phase $V>V_{c}$. The SDW phase looks very similar to the integrable case in \Fig{fig:HUB_excitations}, while in the CDW case both spin and charge excitations are gapped and there is a multitude of additional isolated elementary (or bound state) branches. In the SDW phase -- like in the integrable case -- the lowest variational charge energies are only suboptimal approximations of multi-spin-charge scattering states; the charge-charge and spin-charge continua constructed from the variational energies are therefore not exact, but are kept for reference and marked with dashed boundaries. In the CDW phase the lowest excitation branches are isolated and the accuracy of variational energies -- and consequently also of the multi particle continua -- is expected to be excellent.}
 \label{fig:EHUB_excitations1}
\end{figure*}

\begin{figure*}[p]
 \centering
 \includegraphics[width=0.495\linewidth,keepaspectratio=true]{\figpath/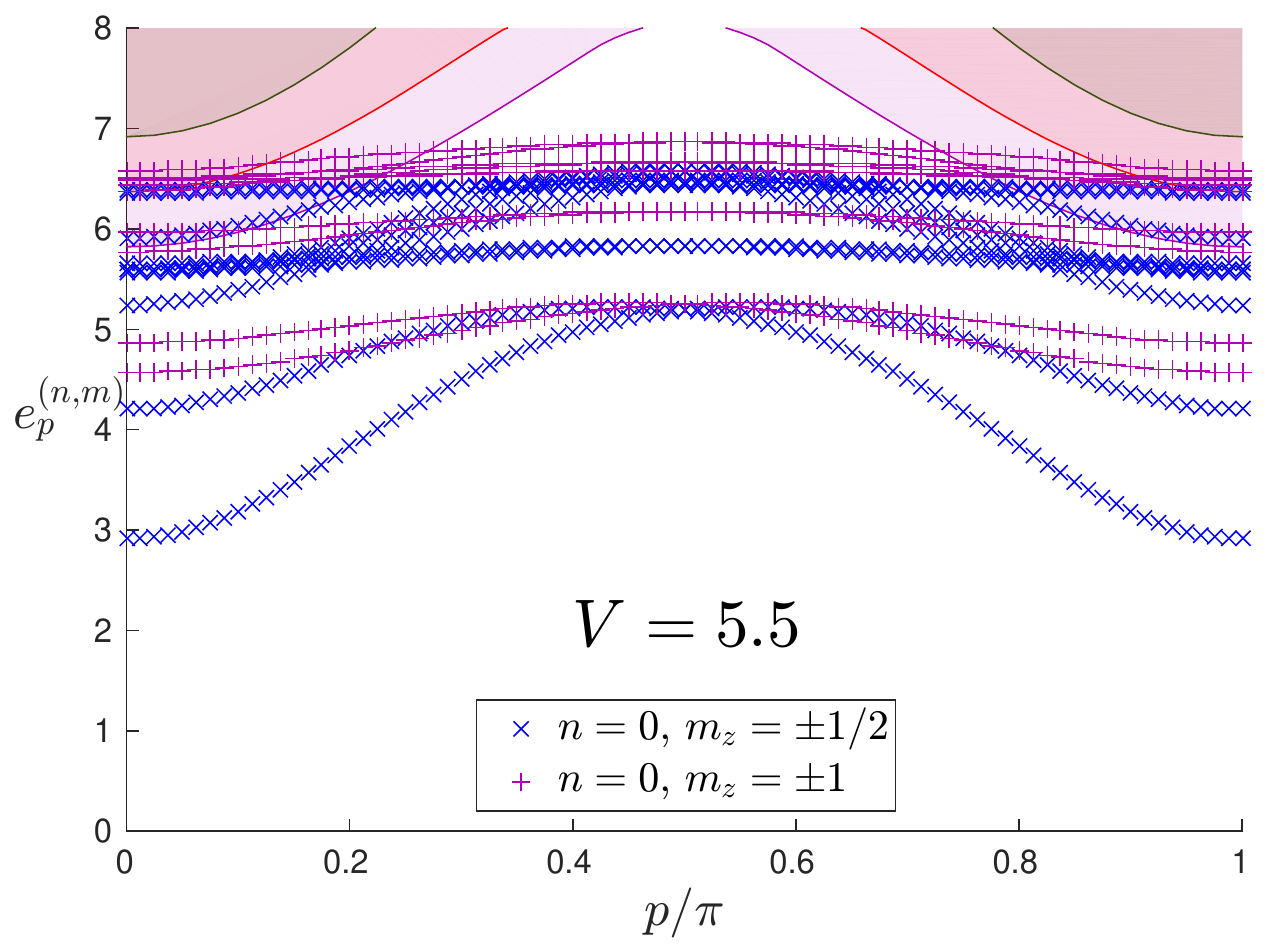}
 \includegraphics[width=0.495\linewidth,keepaspectratio=true]{\figpath/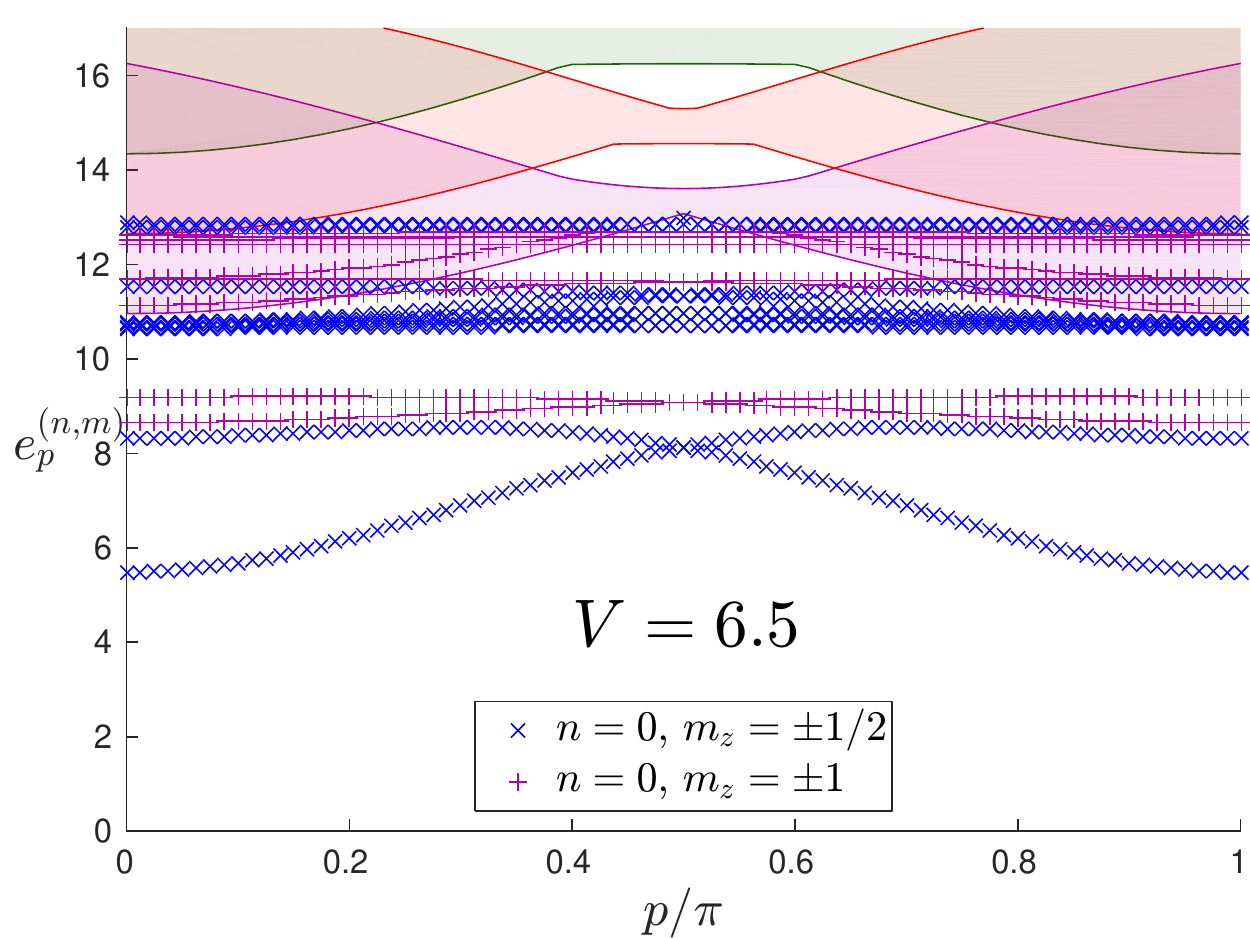}
 \includegraphics[width=0.495\linewidth,keepaspectratio=true]{\figpath/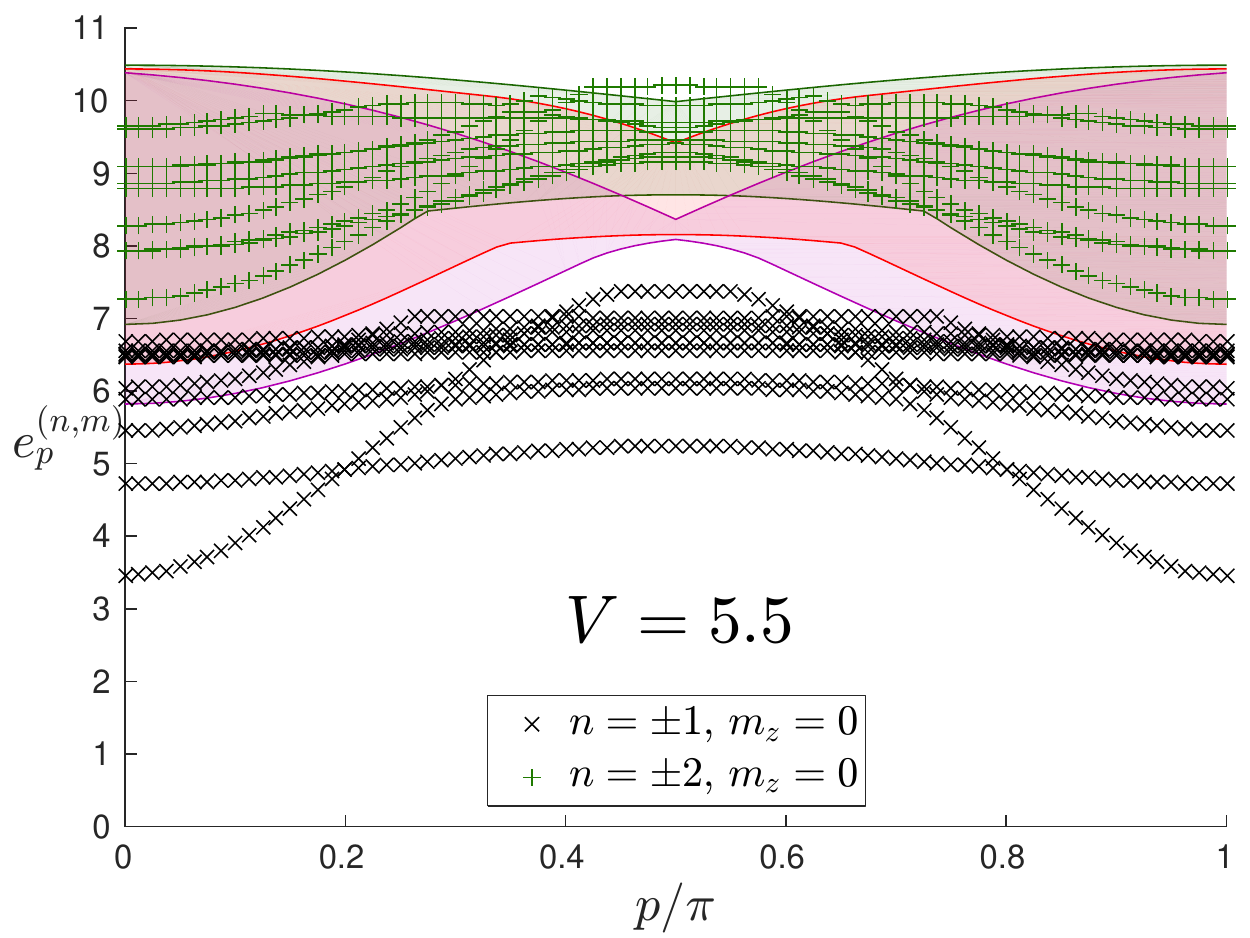}
 \includegraphics[width=0.495\linewidth,keepaspectratio=true]{\figpath/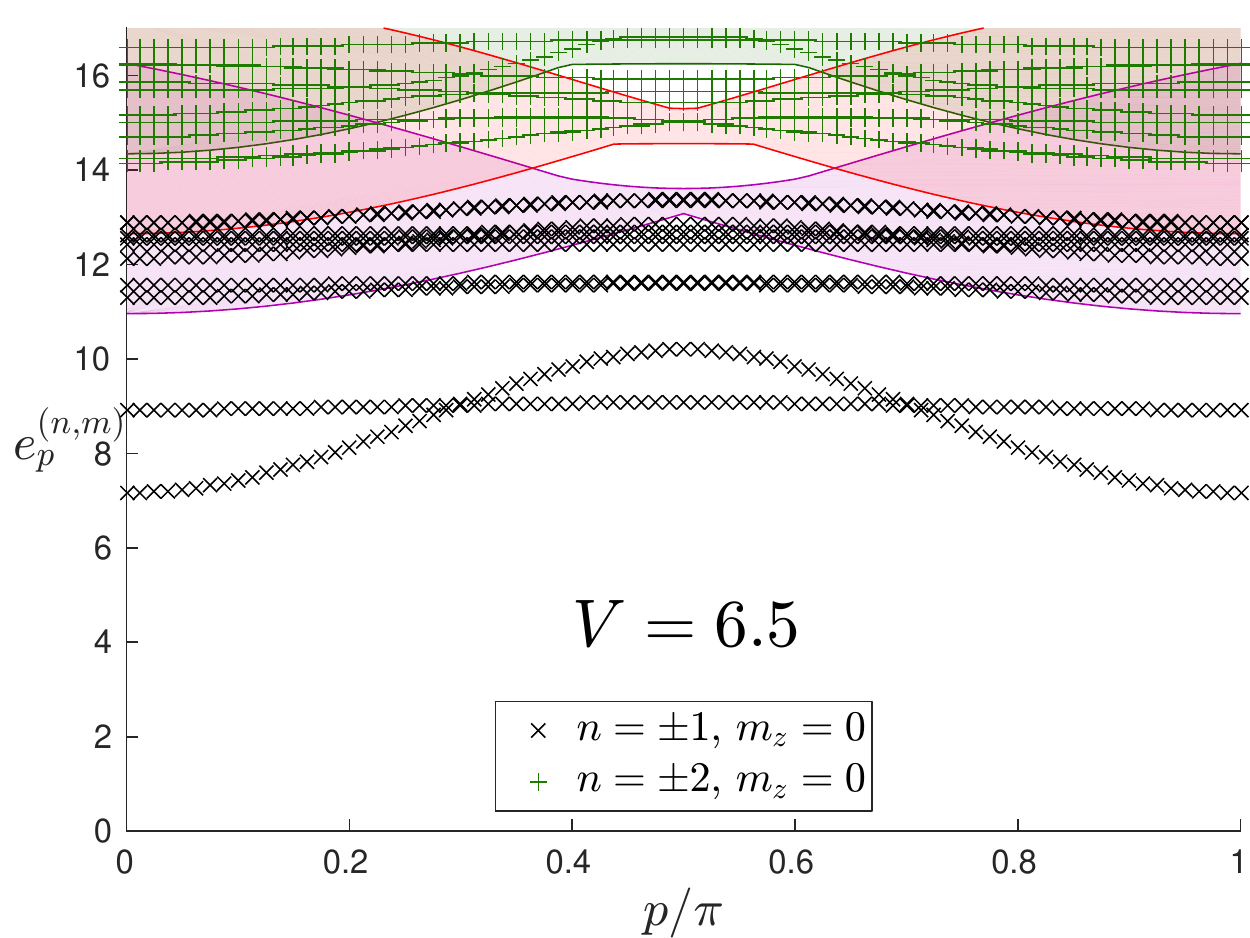}
 \includegraphics[width=0.495\linewidth,keepaspectratio=true]{\figpath/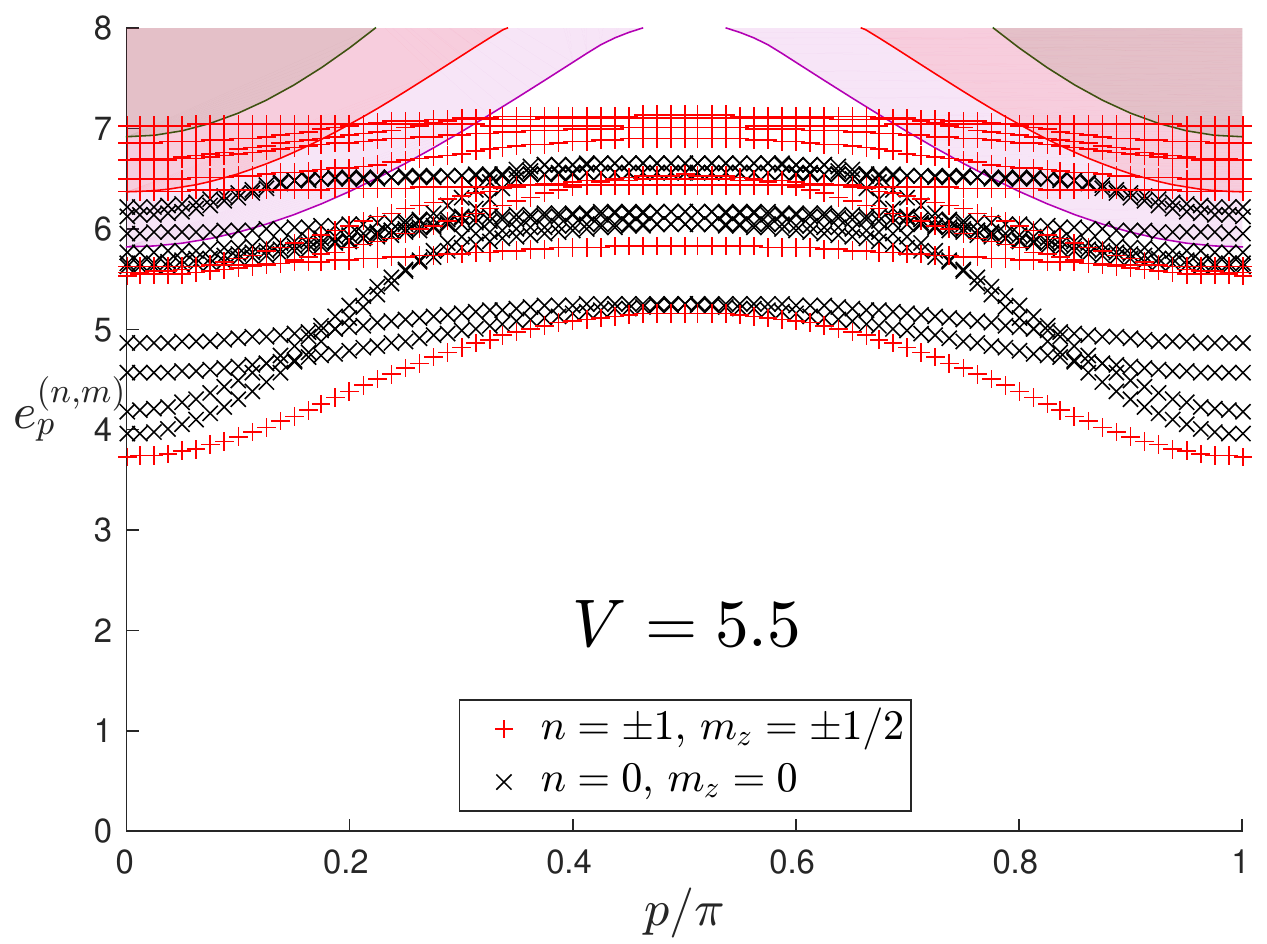}
 \includegraphics[width=0.495\linewidth,keepaspectratio=true]{\figpath/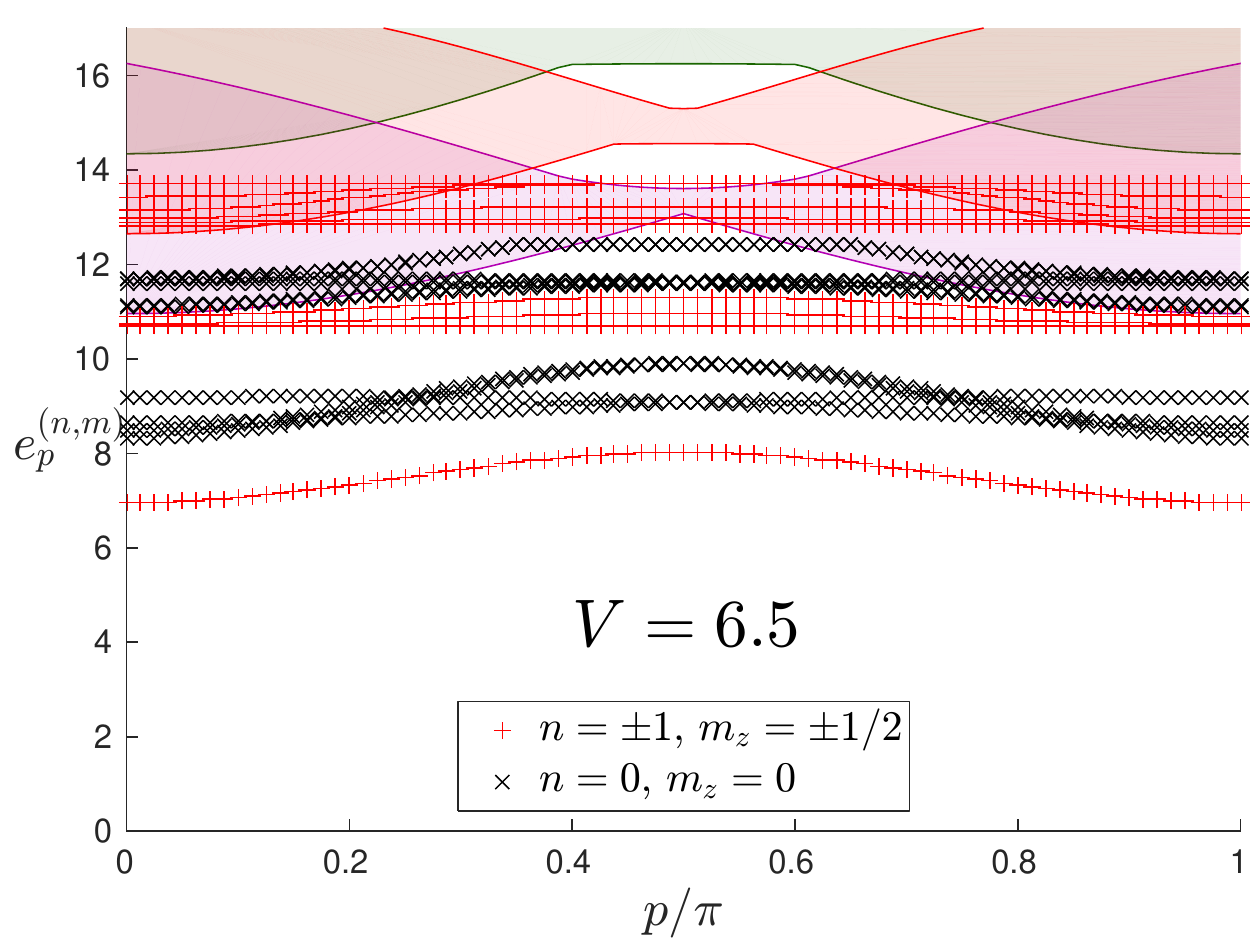}
 \caption{Variational low energy dispersions for various quantum numbers of the extended Fermi Hubbard model \eqref{eq:HUB_Ham} in the non integrable case $V=5.5$ (left) and $V=6.5$ (right) at $U=10$ on top of a half filled ground state ($n_{0}=1$, $m_{0}=0$). We show the first 10 lowest energies for each quantum number, represented by colored symbols. In addition we show spin-spin, charge-charge and spin-charge scattering continua constructed from the variational elementary spin and charge excitations as purple, green and red areas respectively. Here, both spectra are in the CDW phase $V>V_{c}$ (see also caption of \Fig{fig:EHUB_excitations1}).}
 \label{fig:EHUB_excitations2}
\end{figure*}

As a final example, we show the qualitative change of the low energy spectrum of the extended Fermi Hubbard model at half filling at the spin to charge density wave (SDW to CDW) transition for $U,V>0$.\cite{Nakamura00,EjimaNishimoto07} In particular we consider the case of $U=10$ fixed and vary $V$ around the critical point $V_{c}\approx5.13$ of the first order SDW-CDW transition.
 While the charge excitations are always gapped in this parameter regime, the spin excitations are gapless in the SDW phase ($V<V_{c}$) and become gapped in the CDW phase ($V>V_{c}$).

We show results for variational excitation energy dispersions for $V=\{5.1,5.2\}$ and bond dimensions $D=\{400,200\}$ in \Fig{fig:EHUB_excitations1} as well as $V=\{5.5,6.5\}$ and bond dimensions $D=\{120,80\}$ in \Fig{fig:EHUB_excitations2}. We plot the lowest 10 variational energies for various excitation quantum numbers. It can be seen that for $V<V_{c}$ in the SDW phase (top two panels in \Fig{fig:HUB_excitations}) the spin excitations are gapless and the dispersion looks very similar to the integrable case $V=0$ in \Sec{sec:HUB_results}. For $V>V_{c}$ in the CDW phase however, the spin excitations become gapped and the nature of the low energy spectrum changes completely: rather than one single elementary spin and one single elementary charge branch, there is now a multitude of isolated elementary (or bound state) excitation branches which lie below the multi particle scattering continua.

In particular, there are now two lowest excitation branches with spinon quantum numbers $(n=0, m=\pm1/2)$ which have a level crossing at $p=\pi/2$. 
Likewise, there are also several isolated charge excitation branches with holon quantum numbers $(n=\pm1,m=0)$. Here, the elementary holon is now also restricted to half of the first Brillouin zone, as the particle density $n$ shows a strong dimerization in the ground state. While there are several level crossings between these charge branches for $V$ slightly above $V_{c}$, there remain two lowest crossing branches which become more and more separated from the rest with increasing $V$. 

In addition, there are further isolated branches with spinon and holon quantum numbers, and also with electronic, or multi-particle spin or charge quantum numbers, which correspond to bound states. 
These branches lie completely or partially below the spin-spin, charge-charge and spin-charge multi-particle continua constructed from the lowest possible elementary spin or charge excitations (see colored areas in \Fig{fig:EHUB_excitations1} and \Fig{fig:EHUB_excitations2}). For example, for $V\gtrsim 6$ the lowest excitation branch with electronic quantum numbers $(n=\pm1, m=\pm1/2)$ is on the same level as the elementary spin and charge branches and thus far below the spin-charge continuum, and even starts to lie completely below the elementary charge branch. This could be due to a condensation effect that considerably lowers the energy of the spin-charge bound state relative to the individual deconfined elementary spin and charge energies. A more detailed investigation of these effects will be subject of future studies.\cite{CDW_study}

\section{Conclusion}
\label{sec:conclusion}
We present a formulation of the variational MPS ansatz for elementary excitations first proposed in \Ref{MPS_Excitations_variational} with conserved symmetries for multi site unit cells, where the computational cost and the number of variational parameters scales linearly in the number of sites $N$ within the unit cell. The resulting ansatz allows for an efficient separation of the low energy excitation spectrum into certain desired quantum number sectors, which can be targeted individually. This is a great advantage over the original proposal, where an identification of different quantum numbers is only possible a posteriori, and there is no mechanism to target excitations with certain quantum numbers.

We show through the structure of the symmetric ansatz, that elementary excitations in the antiferromagnetic XXZ model (spinons) and in the Fermi Hubbard Model (spinons and holons) are necessarily of topologically nontrivial domain wall nature. Even though such excitations can only exist in pairs in systems with periodic boundary conditions, or be created in pairs locally in an experiment, they are still the theoretical elementary building blocks of all excitations in these systems (analogous to various elementary particles in high energy physics). Even in non-integrable systems however, a variational calculation of such excitations then allows in principle for a systematic construction of the entire spectrum bottom up from these elementary excitations and their scattering behavior.\cite{Laurens_Scattering_2}

The performance of the proposed ansatz is demonstrated by calculating variational low energy dispersions with different quantum numbers for the antiferromagnetic XXZ model and the (extended) Fermi Hubbard model. In cases where exact Bethe ansatz solutions exist for comparison, the elementary spinon excitations are reproduced by the variational ansatz to excellent precision. In the gapped CDW phase of the (non-integrable) extended Fermi Hubbard model, we observe a large number of new bound states below the multi particle continua, which are not present in the gapless SDW phase. It would be interesting to explore the physical consequences of their appearance.

As the gapped elementary holon excitations in the (integrable) Fermi Hubbard model completely lie within a multi particle continuum with the same quantum numbers, the ansatz tries to reproduce lower lying states in this continuum instead, except around momentum $\mom=0,\pi$, where the elementary excitation has the same energy as the lower boundary of the continuum. Possible ways to remedy this fact are discussed in \Sec{sec:HUB_results} and are left to be explored in future studies. Despite this fact, the charge gap can however still be calculated with excellent precision of the same order as the ground state.

We further show that the symmetric ansatz can be used to calculate e.g. the magnetic field $h$ required for a certain ground state magnetization $m_{0}$ of the antiferromagnetic XXZ model. As this strategy allows a direct calculation of $h$, involving a single variational ground state and two (or few) variational excitation calculations only, it is particularly useful for non-integrable models, where otherwise a large number of ground state calculations in a grid search with small variations of $h$ are necessary. This procedure is applicable to all Hamiltonians, which contain generators of their global symmetries as parameterized terms.

The presented ansatz is also a perfect candidate for a more precise and efficient study of elementary excitations in two dimensional systems with topological order on cylinders, such as e.g. in Refs.~\onlinecite{KitaevHB,KagomeExcitations}. More generally, the presented ansatz may prove to be vital for an efficient study of elementary excitations in lattice gauge theories, topological excitations on top of Projected Entangled Pair State (PEPS) \cite{PEPS, MPS5_VMC} ground states in two dimensions, and also for excitations of transfer matrices constructed from topological PEPS.

\section*{Acknowledgments}
We thank 
F. Essler, M. Ganahl, V. Korepin and G. Roose for inspiring and insightful discussions. This work is supported by an Odysseus grant from the FWO, ERC grants QUTE (647905) and ERQUAF (715861), and the EU grant SIQS. V.Z.-S. and F.V. gratefully acknowledge support from the Austrian Science Fund (FWF): F4104 SFB ViCoM and F4014 SFB FoQuS. I.P.M. acknowledges support from the Australian Research Council (ARC) Centre of Excellence for Engineered Quantum Systems, Grant No. CE110001013, and the ARC Future Fellowships Scheme No. FT140100625. The computational results presented have been achieved using the Vienna Scientific Cluster (VSC).

%


\appendix
\section{Application of effective Hamiltonian}
\label{sec:Heff}
In this appendix we describe the necessary terms for applying the effective Hamiltonian $\mathcal{H}^{\rm eff}_{\mom}$ onto a vector $\vec{x}$ of variational parameters, required for solving the effective eigenvalue problem \eqref{eq:eff_EV} using an efficient iterative eigensolver. We restrict to the case of nearest neighbor interactions, i.e. the total Hamiltonian is a sum of nearest neighbor terms $H=\sum_{n}h_{n,n+1}$. The treatment of long ranged Hamiltonians is straight forward to derive, but results in a dramatic increase of the amount and complexity of involved terms. A complete treatment for general Hamiltonians given in terms of Matrix Product Operators (MPOs) will be given elsewhere.

\subsection{Single Site Unit Cells}
\label{sec:Heff_ssuc}
This case has already been considered in the original works and we refer the reader to Refs.~\onlinecite{MPS_Excitations_variational,MPSTP}.

\subsection{Multi Site Unit Cells}
\label{sec:Heff_msuc}
We consider the effective eigenvalue problem \eqref{eq:eff_EV} for a \msuc\ ansatz \eqref{eq:msuc_excitation_ansatz}.
For an efficient solution for low lying excited states we use an iterative Krylov subspace eigensolver (such as e.g. Lanczos). For such methods, only the implementation of the action of $\mathcal{H}^{\rm eff}_{\mom}$ onto some current vector $\vec{x}$ is necessary, which we describe in the following.

We divide the result $\vec{x}_{\rm out}=\mathcal{H}^{\rm eff}_{\mom} \vec{x}_{\rm in}$ of one action of the effective Hamiltonian into individual contributions per site $\vec{x}(n)_{\rm out}$, which are computed separately and combined at the end. Furthermore, we describe all terms in the space of $B$ matrices, where $B(n)^{\s}_{\rm in}=V(n)^{\s}_{L}x(n)_{\rm in}$ and $x(n)_{\rm out}=\sum_{\s}V(n)^{\s\dagger}_{L}B(n)^{\s}_{\rm out}$. 

The individual contributions to $\mathcal{H}^{\rm eff}_{\mom}$ can be derived by fixing the position of $B(n)^{\s}_{\rm out}$ and moving the positions of $B(n)^{\s}_{\rm in}$ and the two site Hamiltonian $h_{n,n+1}$ (cf \Ref{MPS_Excitations_variational}).
Due to the gauge choice \eqref{eq:B_left_gauge} however, a good part of these terms are zero, namely those where $B(n)^{\s}_{\rm in}$ is strictly left of $h_{n,n+1}$ and $B(n)^{\s}_{\rm out}$, and those where $B(n)^{\s}_{\rm out}$ is strictly left of $h_{n,n+1}$ and $B(n)^{\s}_{\rm in}$. Below we give the remaining terms, collected and combined for efficient evaluation. 
Note that here the two site Hamiltonian has been offset by the energy density of the ground state (i.e. $h\to h-e_{0}\unity$ with $e_{0}=\braket{\Psi(\mathbbm{A})|h|\Psi(\mathbbm{A})}$) in order to obtain positive energy differences to the ground state as eigenvalues of \eqref{eq:eff_EV}.

We follow the notation of Refs.~\onlinecite{VUMPS} and \onlinecite{MPS_Excitations_variational} and write $\rbra{x}$ and $\rket{x}$ for vectorizations of a $D\times D$ matrix $x$ in the $D^{2}$ dimensional ``double layer'' virtual space, on which (mixed) transfer matrices, such as $T^{A}_{B}=\sum_{\s}\bar{A}^{\s}\otimes B^{\s}$, or operator transfer matrices, such as 
$O^{AB}_{CD}=\sum_{\s\rho\mu\nu}O^{\s\rho}_{\mu\nu}\bar{A}^{\s}\bar{B}^{\rho}\otimes C^{\mu}D^{\nu}$ (with $O^{\s\rho}_{\mu\nu}=\braket{\s\rho|O|\mu\nu}$),
act.

For better readability we 
also raise site indices to superscripts, e.g. $A(n)^{\s}_{L}\to A^{n,\s}_{L}$ and omit the tilde for $A^{n,\s}_{R}$.

Furthermore we write $T^{n}_{L}=T^{A^{n}_{L}}_{A^{n}_{L}}$ for single site and $T_{L}=\prod_{n=1}^{N}T^{n}_{L}$ for unit cell regular transfer matrices (and similarly for $R$). For mixed single site and unit cell transfer matrices we consequently write ${T^{L}_{R}}^{n}=T^{A^{n}_{L}}_{A^{n}_{R}}$ and $T^{L}_{R}=\prod_{n=1}^{N}{T^{L}_{R}}^{n}$ (and similarly for reversed $L$ and $R$). In all expressions it is understood that $N+1\equiv 1$ and $0\equiv N$.

We start by constructing quantities needed for terms, where $B(n)_{\rm in}$ and $B(n)_{\rm out}$ are on the same site. 
\begin{equation}
\begin{split}
 \rbra{h_{L}^{1}}&=\rbra{\unity}h^{A^{N}_{L}A^{1}_{L}}_{A^{N}_{L}A^{1}_{L}}\\
 \rbra{h_{L}^{n}}&=\rbra{h_{L}^{n-1}}T^{n}_{L} + \rbra{\unity}h^{A^{n-1}_{L}A^{n}_{L}}_{A^{n-1}_{L}A^{n}_{L}},\;n>1
\end{split}
\end{equation} 
and
\begin{equation}
\begin{split}
 \rket{h_{R}^{N}}&=h^{A^{N}_{R}A^{1}_{R}}_{A^{N}_{R}A^{1}_{R}}\rket{\unity}\\
 \rket{h_{R}^{n}}&=T^{n}_{R}\rket{h_{R}^{n+1}} + h^{A^{n}_{R}A^{n+1}_{R}}_{A^{n}_{R}A^{n+1}_{R}}\rket{\unity},\;n<N,
\end{split}
\end{equation} 
which collect Hamiltonian contributions within one unit cell. To collect contributions from all other unit cells we define further
\begin{equation}
\begin{split}
 \rbra{H_{L}^{N}}&=\rbra{h^{N}_{L}}[\unity-T_{L}]^{-1}\\
 \rbra{H_{L}^{n}}&=\rbra{H_{L}^{n-1}}T^{n}_{L},\;n<N
\end{split}
\end{equation}
and
\begin{equation}
\begin{split}
 \rket{H_{R}^{1}}&=[\unity-T_{R}]^{-1}\rket{h^{1}_{R}}\\
 \rket{H_{R}^{n}}&=T^{n}_{R}\rket{H_{R}^{n+1}},\;n>1,
\end{split}
\end{equation}
where $[\unity-T_{L}]^{-1}$ and $[\unity-T_{R}]^{-1}$ are to be understood as pseudo-inverses (cf Ref.~\onlinecite{MPS_Excitations_variational} and Appendix D in Ref.~\onlinecite{VUMPS}).
Finally, we collect all  left and right Hamiltonian contributions up to some site $n$ into
\begin{equation}
\begin{split}
 \rbra{\mathbf{H}^{N}_{L}}&=\rbra{H^{N}_{L}}\\
 \rbra{\mathbf{H}^{n}_{L}}&= \rbra{H^{n}_{L}} + \rbra{h^{n}_{L}},\;n<N
\end{split}
\end{equation}
and
 \begin{equation}
\begin{split}
 \rket{\mathbf{H}^{1}_{R}}&=\rket{H^{1}_{R}}\\
 \rket{\mathbf{H}^{n}_{R}}&= \rket{H^{n}_{R}} + \rket{h^{n}_{R}},\;n>1
\end{split}
\end{equation}
These quantities are independent of $B_{\rm in}$ and $\mom$ and can be precomputed as constants. They will also show up in other subsequent contributions.

We now turn to quantities dependent on $B_{\rm in}$ which have to be recalculated every time $\mathcal{H}^{\rm eff}_{\mom} \vec{x}_{\rm in}$ is invoked. We start with terms
where $B^{n}_{\rm in}$ is right of $B^{n}_{\rm out}$ from within one unit cell
\begin{equation}
\begin{split}
 \rket{b^{N}_{R}}&=T^{A^{N}_{R}}_{B^{N}_{\rm in}}\rket{\unity}\\
 \rket{b^{n}_{R}}&=T^{R}_{L}\rket{b^{n+1}_{R}}+T^{A^{n}_{R}}_{B^{n}_{\rm in}}\rket{\unity},\;n<N
\end{split}
\end{equation} 
and from all other unit cells
\begin{subequations}
\begin{align}
 \rket{B^{1}_{R}}&=[\unity-\fac T^{L}_{R}]^{-1}\rket{b^{1}_{R}}\label{eq:ABR1}\\
 \rket{B^{n}_{R}}&={T^{R}_{L}}^{n}\rket{B^{n+1}_{R}},\;n>1\label{eq:ABRn}
\end{align}
\end{subequations}
Finally, we again collect all such contributions from $B^{n}_{\rm in}$ right of $B^{n}_{\rm out}$ into
\begin{equation}
\begin{split}
 \rket{\mathbf{B}^{1}_{R}}&=\fac\rket{B^{1}_{R}}\\
 \rket{\mathbf{B}^{n}_{R}}&=\fac\rket{B^{n}_{R}} + \rket{b^{n}_{R}}.
\end{split}
\end{equation} 
These terms will be combined with $\rbra{\mathbf{H}^{n}_{L}}$ in the final contributions to $B^{n}_{\rm out}$.

\begin{figure*}[b]
 \centering
 \includegraphics[width=0.95\linewidth,keepaspectratio=true]{\figpath/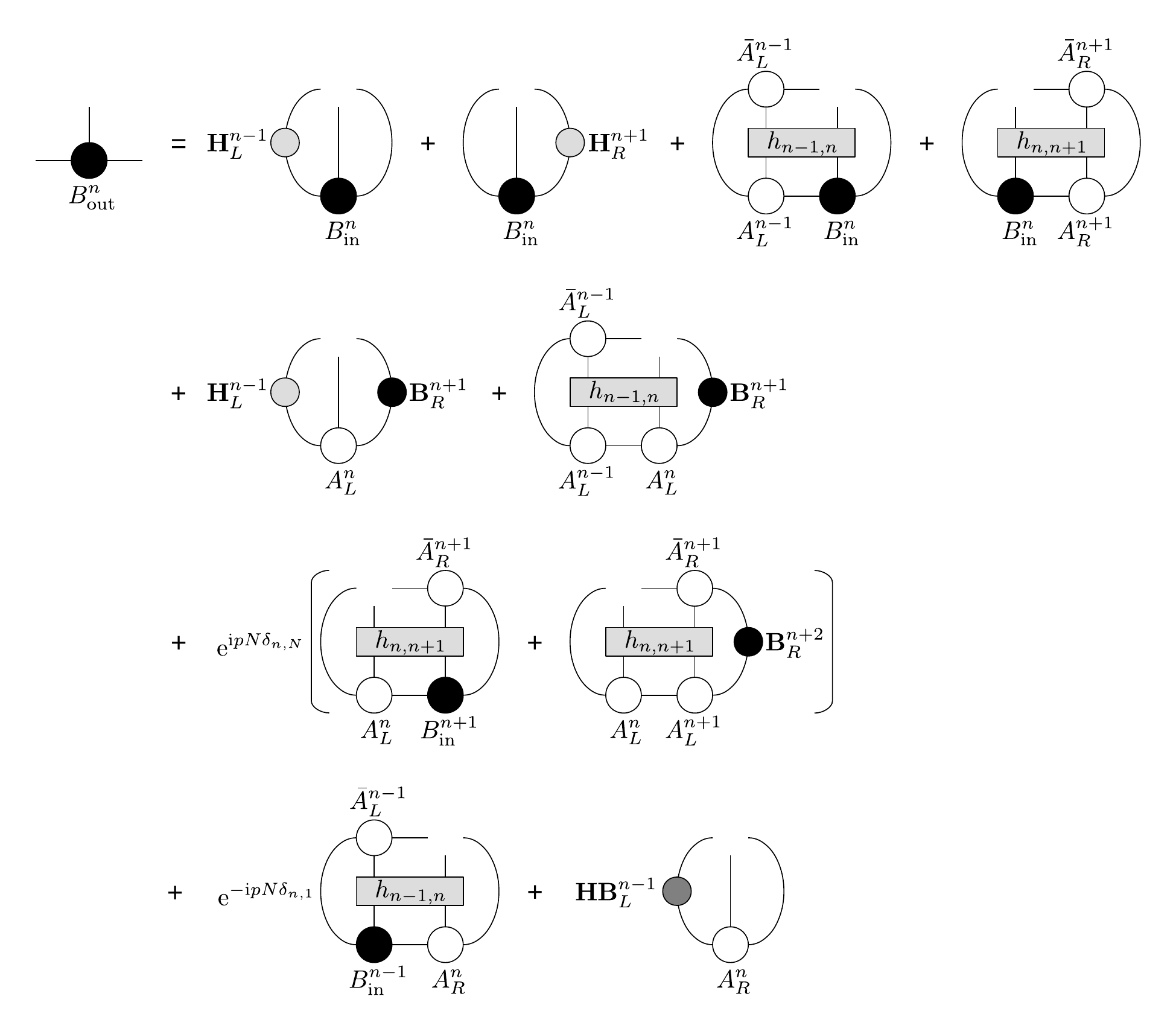}
 \caption{Graphical representation of \eqref{eq:Heff_terms}.}
 \label{fig:Heff_terms}
\end{figure*}

Next we consider quantities where both $h_{n,n+1}$ and $B_{\rm in}$ are left of $B_{\rm out}$
\begin{equation}
\begin{split}
 \rbra{hb_{L}^{1}}=&\rbra{\mathbf{H}^{N}_{L}}T^{A^{1}_{L}}_{B^{1}_{\rm in}} + \rbra{\unity}h^{A^{N}_{L}A^{1}_{L}}_{A^{N}_{L}B^{1}_{\rm in}} + \\
 &\cfac \rbra{\unity}h^{A^{N}_{L}A^{1}_{L}}_{B^{N}_{\rm in}A^{1}_{R}}\\
 \rbra{hb_{L}^{n}}=&\rbra{\mathbf{H}^{n-1}_{L}}T^{A^{n}_{L}}_{B^{n}_{\rm in}} + \rbra{\unity}h^{A^{n-1}_{L}A^{n}_{L}}_{A^{n-1}_{L}B^{n}_{\rm in}} + \\
&\rbra{\unity}h^{A^{n-1}_{L}A^{n}_{L}}_{B^{n-1}_{\rm in}A^{n}_{R}} + \rbra{hb_{L}^{n-1}}{T^{L}_{R}}^{n},\;n>1
\end{split}
\end{equation} 
Here we have collected contributions from $B^{n}_{\rm in}$ within the same unit cell and all $h_{n,n+1}$ left of $B^{n}_{\rm in}$. We proceed to include all contributions of $B^{n}_{\rm in}$ in all other unit cells
\begin{subequations}
\begin{align}
 \rbra{HB_{L}^{N}}&=\rbra{hb_{L}^{N}}[\unity-\cfac T^{L}_{R}]^{-1}\label{eq:HBLN}\\
 \rbra{HB_{L}^{n}}&=\rbra{HB_{L}^{n-1}}{T^{L}_{R}}^{n},\;n<N\label{eq:HBLn}
\end{align}
\end{subequations}
and finally combine
\begin{equation}
\begin{split}
 \rbra{\mathbf{HB}_{L}^{N}}&=\cfac\rbra{HB_{L}^{N}}\\
 \rbra{\mathbf{HB}_{L}^{n}}&=\cfac\rbra{HB_{L}^{n}} +  \rbra{hb_{L}^{n}},\;n<N.
\end{split}
\end{equation} 
The inverses in \eqref{eq:ABR1} and \eqref{eq:HBLN} are to be understood as pseudo-inverses in the case $\mom=0$ and $\xqn=0$ only (i.e. for excitations with zero momentum and the same quantum number(s) as the ground state) and can be fully inverted otherwise, as then the spectral radius of the transfer matrices is strictly smaller than 1.
\begin{widetext}
We now have all the necessary quantities to compute $B^{n}_{\rm out}$
\begin{equation}
\begin{split}
B_{\rm out}^{n,\s}&=\mathbf{H}^{n-1}_{L}B^{n,\s}_{\rm in} + B^{n,\s}_{\rm in}\mathbf{H}^{n+1}_{R}
+\sum_{\rho\mu\nu}h^{\rho\s}_{\mu\nu}(A_{L}^{n-1,\rho})^{\dagger}A^{n-1,\mu}_{L}B^{n,\nu}_{\rm in}
+\sum_{\rho\mu\nu}h^{\s\rho}_{\mu\nu}B^{n,\mu}_{\rm in}A^{n+1,\nu}_{R}(A^{n+1,\rho}_{R})^{\dagger}\\
&+\mathbf{H}^{n-1}_{L}A^{n,\s}_{L}\mathbf{B}^{n+1}_{R}+\sum_{\rho\mu\nu}h^{\rho\s}_{\mu\nu}(A^{n-1,\rho}_{L})^{\dagger}A_{L}^{n-1,\mu}A_{L}^{n,\nu}\mathbf{B}^{n+1}_{R}\\
&+\rme^{\rmi\mom N\delta_{n,N}}\sum_{\rho\mu\nu}h^{\s\rho}_{\mu\nu}A_{L}^{n,\mu}(B_{\rm in}^{n+1,\nu} + A_{L}^{n+1,\nu}\mathbf{B}_{R}^{n+2})(A_{R}^{n+1,\rho})^{\dagger}\\
&+ \rme^{-\rmi\mom N\delta_{n,1}}\sum_{\rho\mu\nu}h^{\rho\s}_{\mu\nu}(A_{L}^{n-1,\rho})^{\dagger}B_{\rm in}^{n-1,\mu}A_{R}^{n,\nu}+ \mathbf{HB}^{n-1}_{L}A_{R}^{n,\s} ,
\end{split}
\label{eq:Heff_terms}
\end{equation} 
where the Kronecker symbols $\delta_{n,N}$ and $\delta_{n,1}$ ensure that the corresponding momentum factors only contribute in the cases $n=N$ and $n=1$ respectively.
Here, the first line corresponds to contributions where $B^{n}_{\rm in}$ and $B^{n}_{\rm out}$ are on the same site, the second and third line where $B^{n}_{\rm in}$ is right of $B^{n}_{\rm out}$, and the last line where $B^{n}_{\rm in}$ is left of $B^{n}_{\rm out}$. For a graphical representation see \Fig{fig:Heff_terms}.

\end{widetext}

\section{Magnetization of Topologically Nontrivial Excitations from Translation Symmetry Broken Ground States in the XXZ Model}
\label{sec:m13_excitations}

In this Appendix we elaborate on the topologically nontrivial elementary excitations with fractional magnetizations obtained from the threefold degenerate, translation symmetry broken ground states with magnetization density $m_{0}=1/6$ of \eqref{eq:XXZ_Ham} considered in \Sec{sec:h_from_e}.
As a consequence of the broken translation symmetry, the single site ground state magnetizations $m_{0,n}\neq 1/6$, while $\sum_{n}^{N}(m_{0,n}-1/6)=0$ still holds within one unit cell. Denote the suMPS unit cells for these three degenerate ground states as $\mathbbm{A}_{1}$, $\mathbbm{A}_{2}$ and $\mathbbm{A}_{3}$ and assume $\ket{\Psi(\mathbbm{A}_{3})}=T\ket{\Psi(\mathbbm{A}_{2})}=T^{2}\ket{\Psi(\mathbbm{A}_{1})}$.

The resulting magnetizations of domain wall excitations created from combining these three different ground states are not exactly equal to the quantum numbers $\xqn$. This is due to the fact, that the perturbation unit cell $\mathbbm{B}$ consists of a superposition of different contributions from $\mathbbm{A}_{i,L}$ and $\mathbbm{A}_{j,R}$ ($i\neq j$), such that now $\sum_{n}(m_{0,n}- m_{0})\neq 0$ within the perturbation unit cell. Consequently, $\mathbbm{B}$ then has a magnetization different from $Nm_{0}$ and the excitation carries a magnetization slightly perturbed away from $\xqn$. These perturbations are usually of the form $m_{0,n}-m_{0}$. For example, if we take the $m_{0,n}$ to be the magnetizations of the $\mathbbm{A}_{i,L}$ unit cell, the $\xqn=1/3$ excitation then carries effective magnetization $m=1/3-(m_{0,3}-1/6)$, while the $\xqn=-1/3$ excitation carries $m=-1/3+(m_{0,1}-1/6)$.

However, the three possible excitations for each $\xqn$ (which are related by single or two site overall translations of the entire state) together have again a mean magnetization of exactly $m=\xqn$. For example, a $\xqn=1/3$ excitations can be generated from combining $\mathbbm{A}_{1,L}$ with $\mathbbm{A}_{2,R}$, $\mathbbm{A}_{2,L}$ with $\mathbbm{A}_{3,R}$ or $\mathbbm{A}_{3,L}$ with $\mathbbm{A}_{1,R}$.
To remedy the above fact -- which is once more an artifact of open boundary conditions and finite bond dimension -- we therefore compute and average over all three excitation energies for each $\xqn$. Exactly this has been done to obtain the values shown in \Tab{tab:eq_m13}.

Note that for topologically \textit{trivial} excitations the magnetization is always well defined and precisely corresponds to $m=\xqn$. This is because the same MPS ground state unit cell is used left and right of the perturbation matrix $\mathbbm{B}$.

\end{document}